\documentclass{aa} 
\usepackage{graphicx}
\usepackage{txfonts}
\usepackage{tabularx}
\usepackage{appendix}
\usepackage{threeparttable}  
\usepackage{dcolumn}
\usepackage{amsmath}

\newcommand{\hii}{H{\sc\,ii}}
\newcommand{\uchii}{UC\,H{\sc\,ii}}
\newcommand{\hchii}{HC\,H{\sc\,ii}}

\newcommand{\kms}{\,km\,s$^{-1}$} 

\newcommand{\coa}{$^{13}$CO\,(2\,--\,1)}
\newcommand{\cob}{C$^{18}$O\,(2\,--\,1)}
\newcommand{\coaa}{$^{13}$CO}
\newcommand{\cobb}{C$^{18}$O}

\newcommand{\coab}{$^{13}$CO\,(3\,--\,2)}
\newcommand{\coaf}{$^{13}$CO\,(2\,--\,1)}

\newcommand{\coac}{CO\,(3\,--\,2)}

\newcommand{\coae}{CO\,(1\,--\,0)}

\newcommand{\msunpc}{$\rm M_\odot\,\rm pc^{-2}$}

\newcolumntype{d}[1]{D{.}{\cdot}{#1}}

\usepackage[normalem]{ulem}

\usepackage[usenames,dvipsnames]{xcolor}

\definecolor{tbc}{cmyk}{0.05002,1,0.9,0}

\definecolor{mygreen}{cmyk}{0.85002,0.3,1,0}

\definecolor{myyellow}{cmyk}{0.00,0.24,0.86,0.20}

\definecolor{mypink}{cmyk}{0.00,0.80,0.00,0.00}

\definecolor{grey}{cmyk}{0.30,0.30,0.30,0.30}

\usepackage{url}

\begin{document} 
\newcolumntype{L}[1]{>{\raggedright\arraybackslash}p{#1}}
\newcolumntype{C}[1]{>{\centering\arraybackslash}p{#1}}
\newcolumntype{R}[1]{>{\raggedleft\arraybackslash}p{#1}}
\title{The SEDIGISM survey: A search for molecular outflows}
\author{A.\,Y.\,Yang 
        \inst{1}\thanks{E-mail: ayyang@mpifr-bonn.mpg.de}, 
       J.\,S.\,Urquhart\inst{2},
         F.\,Wyrowski \inst{1},
         M.\,A.\,Thompson\inst{3},   
        C.\,K{\" o}nig\inst{1},
        D.\,Colombo\inst{1}, 
         K.\,M.\,Menten \inst{1},
         A.\,Duarte-Cabral\inst{4},
         F.\,Schuller\inst{1,5},
         T.\,Csengeri\inst{6},
         D.\,Eden\inst{7}, 
         P.\,Barnes\inst{8,16}, 
         A.\,Traficante\inst{9},
         L.\,Bronfman\inst{10},
         A.\,Sanchez-Monge\inst{11},
         A.\,Ginsburg\inst{12},
         R.\,Cesaroni\inst{14},
         M.-Y.\,Lee\inst{13},
        H.\,Beuther\inst{15},
        S.-N.\,X.\,Medina\inst{1},
        P.\,Mazumdar\inst{1},
        T.\,Henning\inst{15} }
\institute{\textit{Affiliations are listed in the end of the paper}}

   \date{Received 17/08/2021; accepted: 19/11/2021  }
%
  \abstract
   { 
   The formation processes of massive stars are still unclear, but a picture is emerging involving accretion disks and molecular outflows in what appears to be a scaled-up version of low-mass star formation. 
   A census of outflow activity toward high-mass star-forming clumps in various evolutionary stages has the potential to shed light on high-mass star formation. 
   }
   {
   We conducted an outflow survey toward ATLASGAL (APEX Telescope Large Area Survey of the Galaxy) clumps using SEDIGISM (structure, Excitation, and Dynamics of the Inner Galactic InterStellar Medium) data and aimed to obtain a large sample of clumps exhibiting outflow activity in different evolutionary stages. } 
   {  
   We identify the high-velocity wings of the \coaa\ lines, which indicate outflow activity, toward ATLASGAL clumps by (1) extracting the simultaneously observed  \coa\ and \cob\ spectra from SEDIGISM, and (2) subtracting Gaussian fits to the scaled \cobb\ (core emission) from the \coaa\, line after considering opacity broadening. 
 
   }
   { 
   We detected high-velocity gas toward 1192 clumps out of a total sample of 2052,  corresponding to an overall detection rate of 58\%. 
   Outflow activity has been detected in the earliest (apparently) quiescent clumps (i.e., $70\rm \mu m$ weak) to the most evolved \hii\, region stages (i.e., $8\rm \mu m$ bright with tracers of massive star formation).
    The detection rate increases as a function of evolution (quiescent=51\%, protostellar=47\%, YSO = 57\%, \uchii\ regions = 76\%).  
   } 
   { 
    Our sample is the largest outflow sample identified so far. 
   The high detection rate from this large sample is consistent with the results of similar studies reported in the literature and supports the  scenario that outflows are a ubiquitous feature of high-mass star formation. 
  The lower detection rate in early evolutionary stages may be due to the fact that outflows in the early stages are weak and difficult to detect.
  We obtain a statistically significant sample of outflow clumps for every evolutionary stage, especially for outflow clumps in the earliest stage (i.e., 70\,$\mu$m dark).
  The detections of outflows in the 70\,$\mu$m dark clumps suggest that the absence of 70\,$\mu$m emission is not a robust indicator of starless and/or pre-stellar cores. 
  }
\titlerunning{A search for Outflow Clumps}
\authorrunning{Yang et al.}
\maketitle
%
\section{Introduction}
 
The formation processes of massive stars within molecular clumps are still unclear \citep{Zinnecker2007ARAA481Z,Krumholz2019ARAA57227K,Motte2018ARAA5641M}.
However, progress is being made in understanding the accretion mechanism. 
With the growing observational evidence of disk-like structures around OB-type protostars \citep[e.g.,][]{Kuiper2011ApJ73220K,Beuther2012AA543A88B,Boley2013AA558A24B,Haemmerle2016AA585A65H,Ilee2018ApJ869L24I,Csengeri2018AA617A89C,Maud2019AA627L6M,Beuther2019AA621A122B,Zapata2019ApJ872176Z,Goddi2020ApJ90525G,Zhao2020SSRv21643Z}, the picture and the role of accretion disks in the formation of massive stars is becoming increasingly clear, supporting the scenario that high-mass star formation is a scaled-up version of low-mass star formation involving disk accretion and molecular outflows \citep[e.g.,][]{Beuther2002AA383B,Duarte-Cabral2013AA558A125D,Beltran2016AARv246B,Ilee2018ApJ869L24I}.
Direct observational evidence of disks around OB-type protostars, though, is limited, and many details remain uncertain \citep{Beltran2016AARv246B,Goddi2020ApJ90525G}. 

The small-scale disks around massive stars in complex embedded environments can be indirectly inferred by: (1) the presence of highly collimated molecular outflows \citep{Goddi2020ApJ90525G}, (2) the existence of velocity gradients perpendicular to outflows \citep{Ginsburg2018ApJ860119G,Maud2018AA620A31M}, (3) elongated compact emission perpendicular to outflow orientation \citep{Kraus2010Natur466339K}, and (4) 
the presence of outflow direction perpendicular to the distribution of the methanol and/or water masers \citep[e.g.,][]{Beltran2016AARv246B}. 
In particular, large-scale outflows are found to be perpendicular to the small-scale disk around massive protostars \citep{Kraus2010Natur466339K,Beltran2016AARv246B}. 
Molecular outflows are, therefore, a crucial first step for selecting good candidates with disk-like structures that can then be studied in detail. 

Outflows can be inferred by the presence of high-velocity emission in the wings of molecular lines \citep[e.g.,][]{Snell1980ApJ239L,Zhang2001ApJ552L,Beuther2002AA383B,Arce2007prplconf245A,deVilliers2014MNRAS444,Maud2015MNRAS645M,Yang2018ApJS235Y,Li2018ApJ867167L}, which have been detected in the observational stages of massive star formation as classified by \citet{Zinnecker2007ARAA481Z}, from the earliest infrared dark phase \citep{Beuther2005ApJ634L185B,Duarte-Cabral2013AA558A125D,Feng2016ApJ100F,Tan2016ApJ3T}, to hot cores \citep{Kurtz2000prpl299K}, to the \uchii\ region phase \citep{Codella2004AA615C,Qin2008AA361Q,Yang2018ApJS235Y}. 
Outflows are thus a useful tool for improving our understanding of the accretion process in every stage of massive star formation.

However, either these studies are targeted observations or the sample sizes in the various stages are very small. 
Large outflow surveys are needed to provide statistically significant samples for clumps in different evolutionary stages. 
\citet{Yang2018ApJS235Y} conducted an unbiased outflow survey using data from the CHIMPS (the \coaa/\cobb\,(J=3-2) Heterodyne Inner Milky Way Plane Survey, \citealt{Rigby2016MNRAS456R}) and found that there is an evolutionary trend of the outflow detection rate to increase as clumps evolve, and suggested that clump-scale outflows are dominated by the most massive and luminous source within the clump. 
Only two sources associated with outflows were determined to be in the earliest stages (i.e., 70$\mu m$ dark sources). 
These 70$\mu m$ dark sources are thought to be massive, starless, and early-stage core candidates \citep[e.g.,][]{Feng2016ApJ100F,Tan2016ApJ3T}, but the associated outflows imply that there is star formation within them. These outflows could identify clumps in the earliest stages of star formation before they become mid-infrared visible, and to obtain a statistically significant sample of these clumps in the earliest stages, large surveys of outflows are required. 
Detailed, high-resolution interferometric observations toward these young outflow clumps are ultimately required to study the outflows at a sufficient resolution to pinpoint their origin and to understand the complex processes involved. 

In this paper, we present the largest survey of CO outflows carried out to date by combining the ATLASGAL (APEX Telescope Large Area Survey of the Galaxy) and SEDIGISM (Structure, Excitation, and Dynamics of the Inner Galactic InterStellar Medium)  surveys. 
This large observational sample of outflows provides statistically significant and well-selected subsamples across a range of evolutionary stages of clumps.
This work is a continuation of a series of outflow studies building on and extending the results presented in \citet[][hereafter Paper\,I]{Yang2018ApJS235Y}, who conducted an outflow survey and statistical analysis of the outflow properties for clumps in different evolutionary stages. This paper infers the presence of outflows due to the observation of line wings and discusses the detection statistics of outflows.
In Sect.\,\ref{sect:surveys} we provide an overview of the surveys used and the sample selection.
We describe the outflow wing identification method in Sect.\,\ref{sect:wings_identification} and present the results of outflow detection and the statistics of detection rates in Sect.\,\ref{sect:results}. 
In Sect.\,\ref{sect:discussion}, we discuss the potential implications of our results for the high-mass star-formation process and we summarize our work and present our conclusions in Sect.\,\ref{sect:summary_conclusion}. 

\begin{figure*}
\centering
\includegraphics[width = 0.89\textwidth] {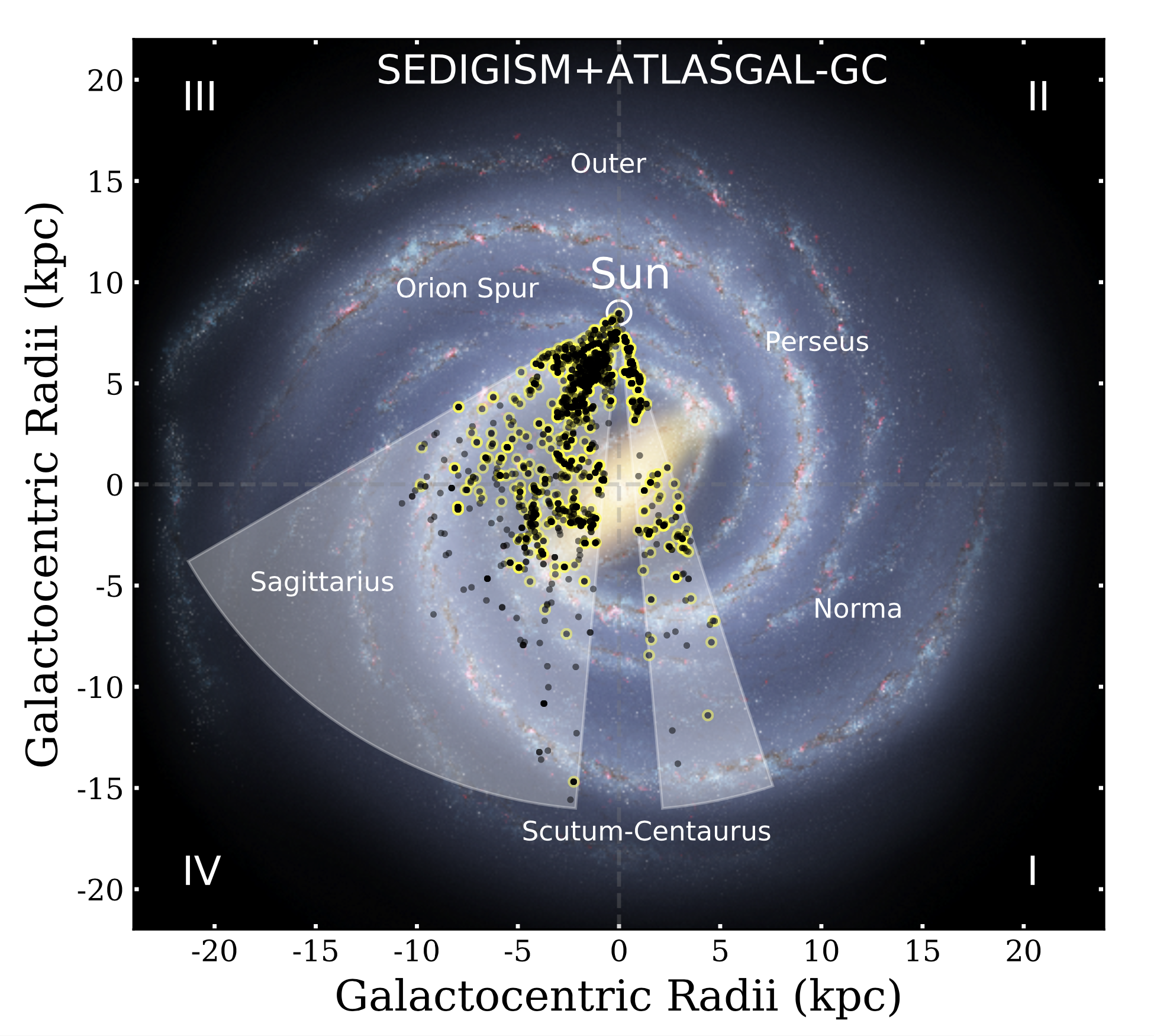}
 \caption{A Figure of the Galaxy showing the region covered by this work. The background image is an artist's impression of the large structure of the Galactic plane as viewed from the Northern Galactic Pole [courtesy of NASA/JPL-Caltech/R. Hurt (SSC/Caltech)]. 
 The white shaded area indicates the covered region by this work (SEDIGISM+ATLASGAL-GC) with the Galactic Center (GC) region\,($\vert \ell \vert <5\degr$) excluded. 
 The white symbol $\odot$ indicates the position of the Sun, and the black dots and yellow circles indicate the positions of the total 4120 and the selected 2052 clumps, respectively. 
The spiral arms are labeled in white and the Galactic quadrants are shown in the corners as Roman numerals. 
 }
\label{fig:galactic_distribution}
\end{figure*}

\setlength{\tabcolsep}{6pt}
\begin{table*}
\centering
\caption[]{ \it \rm Clump properties of the selected  2052 ATLASGAL clumps  searched for outflows: clumps Galactic name, heliocentric distance (Dist.), bolometric luminosity ($\rm L_{bol}$), clump mass ($\rm M_{clump}$), the peak $\rm H_{2}$ column density [$\rm N({H_2})$], mean volume density ($\rm n_{H_2}$) and mean mass surface density ($\Sigma$) at FWHM level (i.e., within the 50\% contour), the associations with methanol maser ($\rm CH_{3}OH$), water maser ($\rm H_{2}O$), and SiO emission. }
\begin{tabular}{llllllllll}
\hline
\hline
ATLASGAL  & Dist. & $\rm \log L_{bol}$ & $\rm \log M_{clump}$ & $\rm \log N({H_2})$   & $\rm \log n_{H_2}$ &  $\rm \ log \Sigma$  &  $\rm CH_{3}OH$ & $\rm H_{2}O$ & SiO \\
(CSC Gname   ) & (kpc) & ($\rm L_\odot$)  &  ($\rm M_{\odot}$)  & ($\rm cm^{-2} $)  & ($\rm cm^{-3}$) & ($\rm M_{\odot}\,pc^{-2}$) & $-$ & $-$  &$-$ \\

\hline
AGAL005.001$+$00.086 & 2.93 & 2.29 & 2.07 & 22.17 & 4.59 & 2.89 & N & N & N \\
AGAL005.076$-$00.091 & 10.80 & 4.02 & 3.64 & 22.36 & 4.2 & 3.16 & N & N & N \\
AGAL005.321$+$00.184 & 2.94 & 1.92 & 2.18 & 22.08 &$-$&$-$& N & N & N \\
AGAL005.371$+$00.319 & 12.81 & 4.04 & 2.8 & 22.06 & 4.03 & 2.76 & N & N & N \\
AGAL005.387$+$00.187 & 2.94 & 1.66 & 2.3 & 22.71 & 5.15 & 3.34 & N & N & N \\
AGAL005.397$+$00.194 & 2.94 & 1.96 & 2.59 & 22.62 & 4.66 & 3.11 & N & N & N \\
AGAL005.474$-$00.244 & 2.96 & 4.22 & 2.83 & 22.48 & 4.2 & 2.89 & N & N & N \\
$\vdots$  & $\vdots$   &  $\vdots$  & $\vdots$   & $\vdots$   & $\vdots$   & $\vdots$   & $\vdots$   & $\vdots$   &  $\vdots$ \\
AGAL005.884$-$00.392 & 3.0 & 5.33 & 2.72 & 23.49 & 6.14 & 4.14 & Y & Y & Y \\
AGAL005.897$-$00.444 & 3.0 & 4.88 & 2.8 & 22.93 & 4.83 & 3.3 & N & Y & N \\
AGAL005.899$-$00.429 & 3.0 & 4.75 & 2.71 & 23.3 & 5.73 & 3.86 & Y & Y & N \\
\hline
\hline
\end{tabular}
\tablefoot{These physical properties are taken from \citet{Urquhart2018MNRAS4731059U} and \citet{Urquhart2021arXiv211112816U}, with uncertainties of a factor of a few.  
 The information of maser associations are from \citet{Billington2020MNRAS4992744B},  with the detection levels of  15-167\,mJy (1$\sigma$) for $\rm H_{2}O$ \citep {Walsh2011MNRAS1764W,Walsh2014MNRAS2240W} and $\sim$170\,mJy (1$\sigma$) for $\rm CH_{3}OH$ maser \citep{Green2009MNRAS783G}. 
 The SiO data are collected from the SEDIGISM survey and the literature \citep{Harju1998AAS132211H,Csengeri2016AA586A149C,Stroh2019ApJS24425S}, with sensitivities of $\sim 0.8$\,K \citep{Schuller2017AA601A124S}, $\sim$0.03-0.07\,K  \citep{Harju1998AAS132211H}, 15-30\,mK  \citep{Csengeri2016AA586A149C}, and 14 mJy/beam \citep{Stroh2019ApJS24425S}.
 The symbol "Y" means maser detections. $-$ means no measurements, whilst "N" means no detections. 
 Only a small part of the table is presented here with the full version available from CDS.}
\label{tab:clump_properties}
\end{table*}

\setlength{\tabcolsep}{3pt}
\begin{figure*}[!htb]
\centering
\begin{tabular}{ccc}
 \includegraphics[width = 0.32\textwidth] {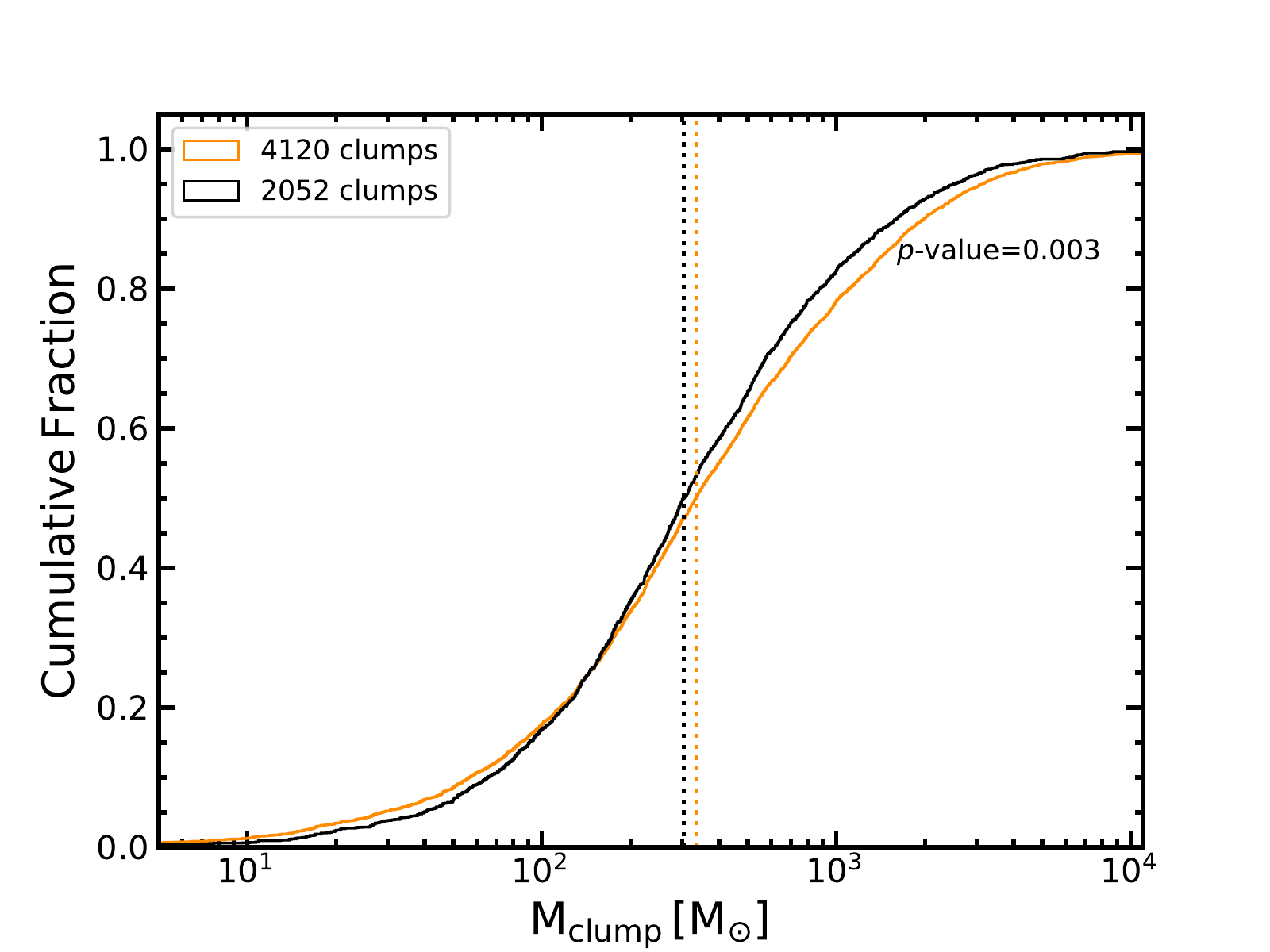} & 
 \includegraphics[width = 0.32\textwidth] 
 {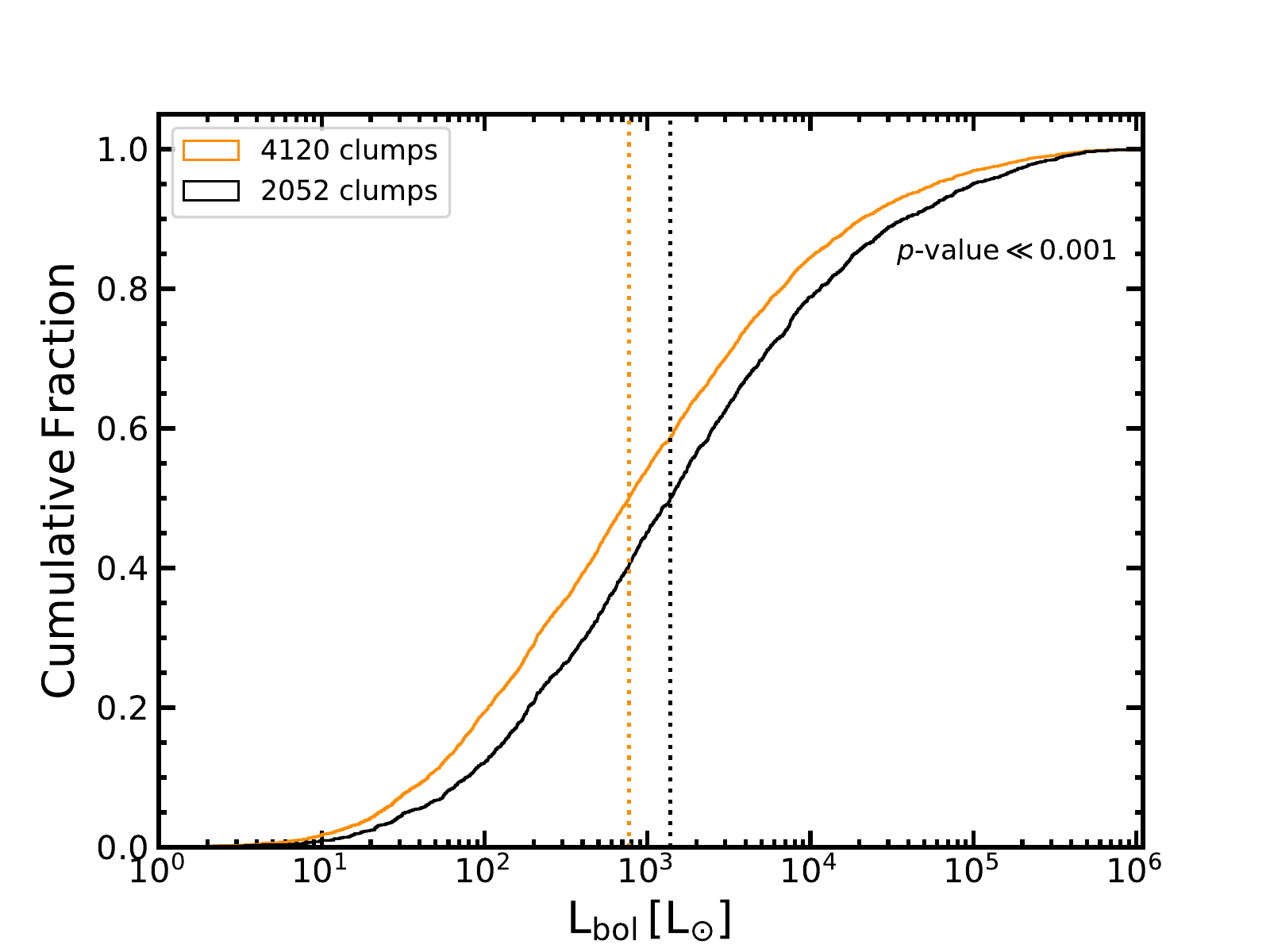} & \includegraphics[width = 0.32\textwidth] 
 {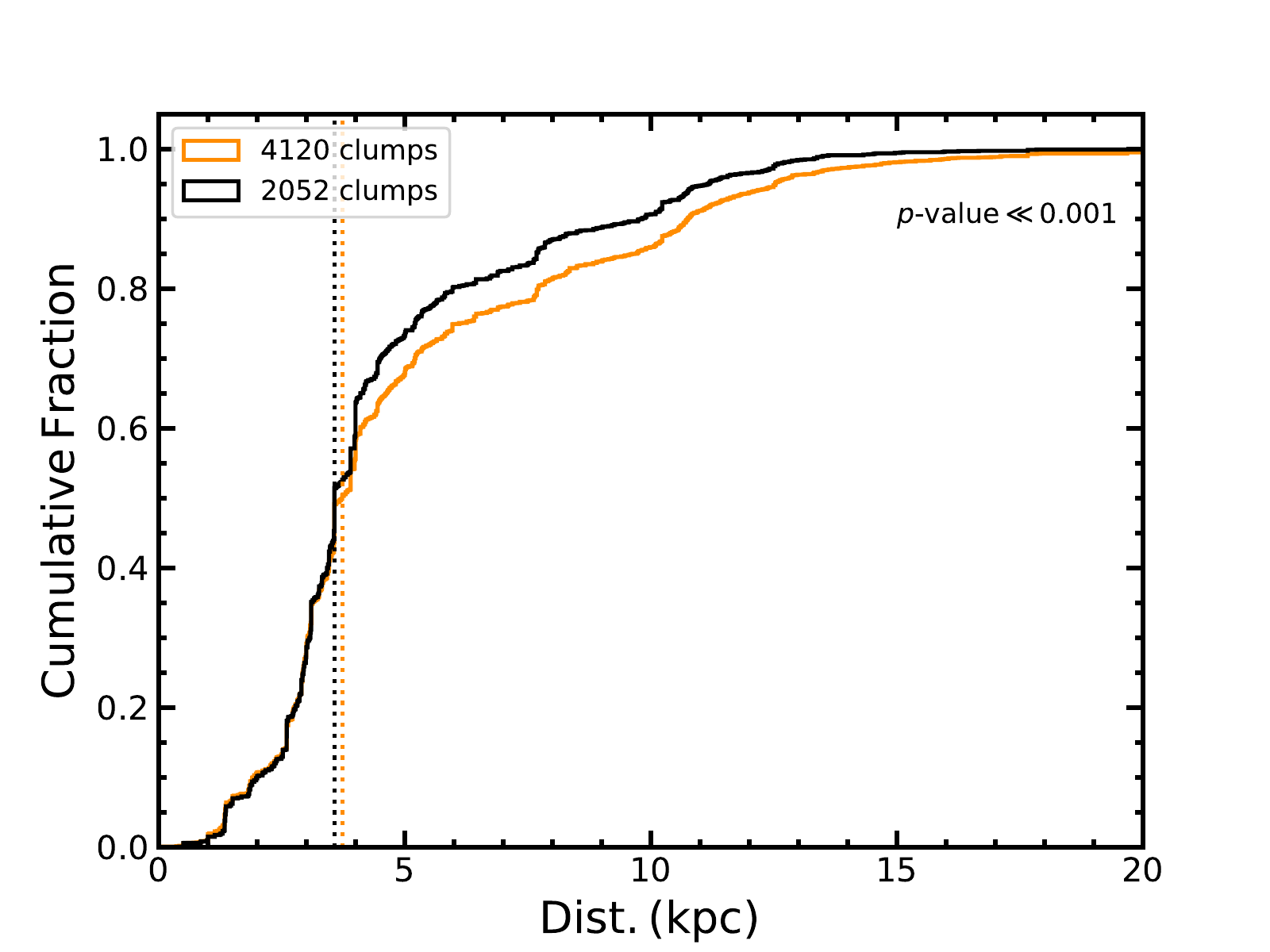} \\
 (a)& (b) & (c) \\
   \includegraphics[width = 0.32\textwidth] {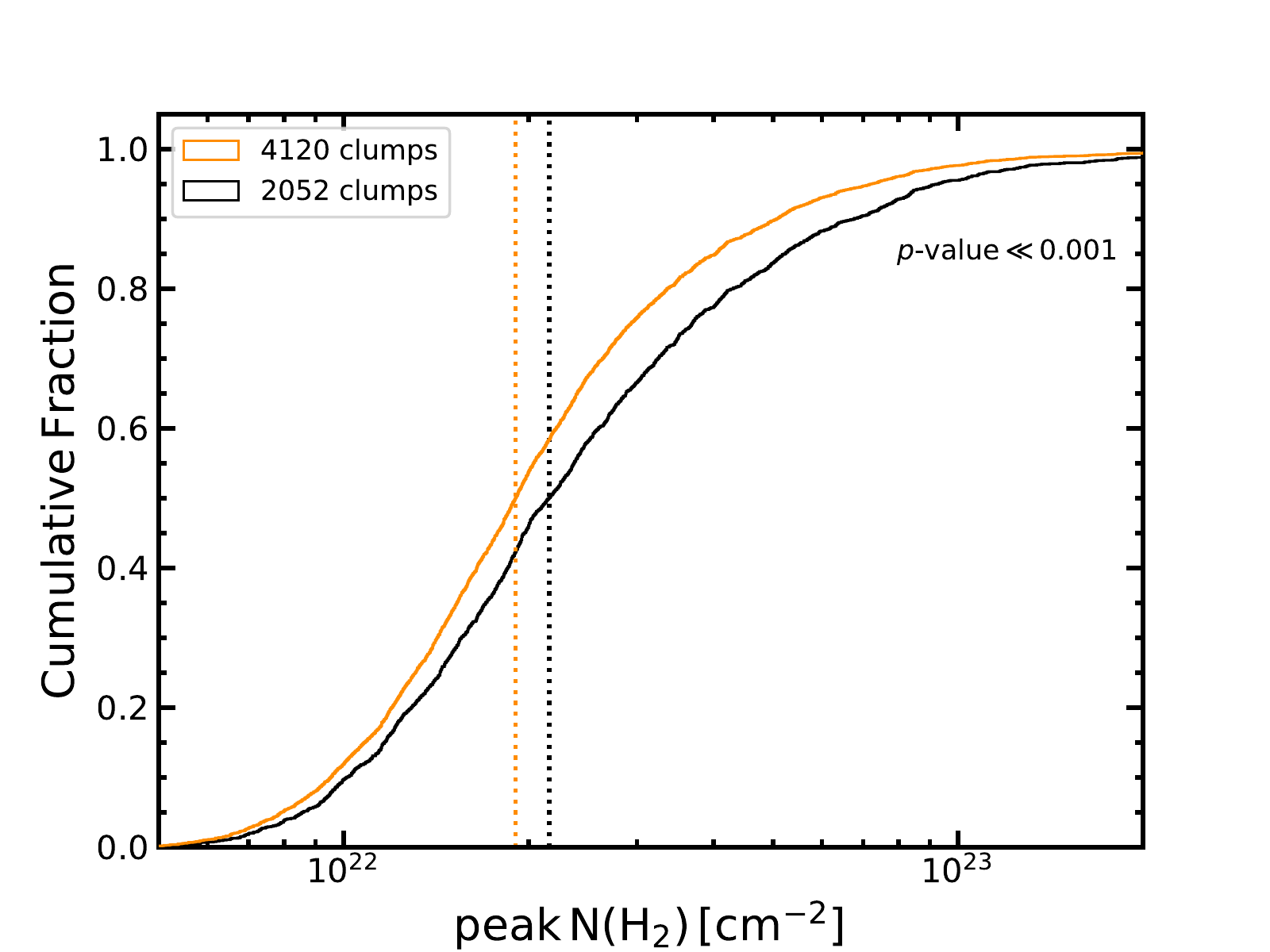} &
  \includegraphics[width = 0.32\textwidth] {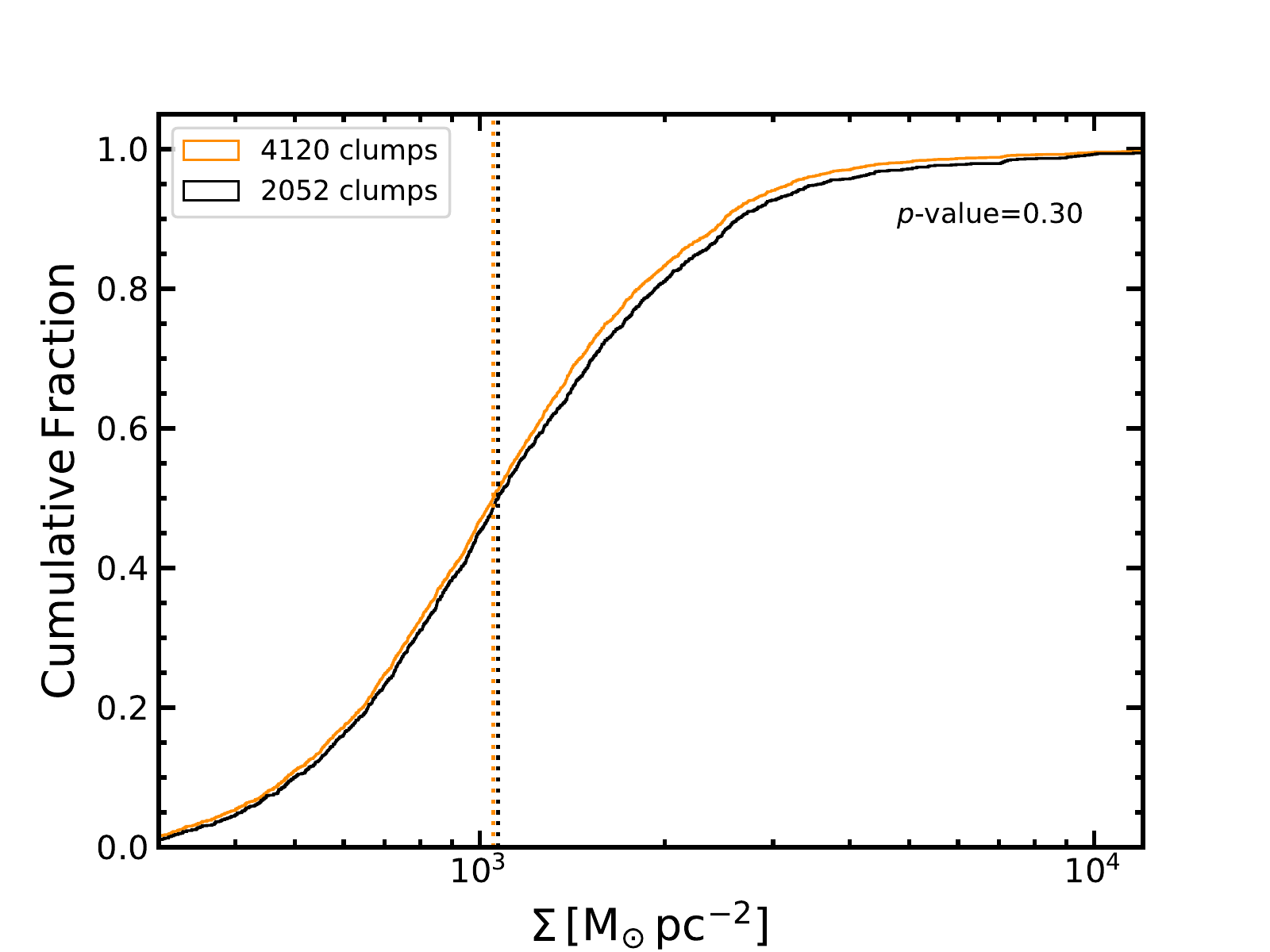} &
 \includegraphics[width = 0.32\textwidth] {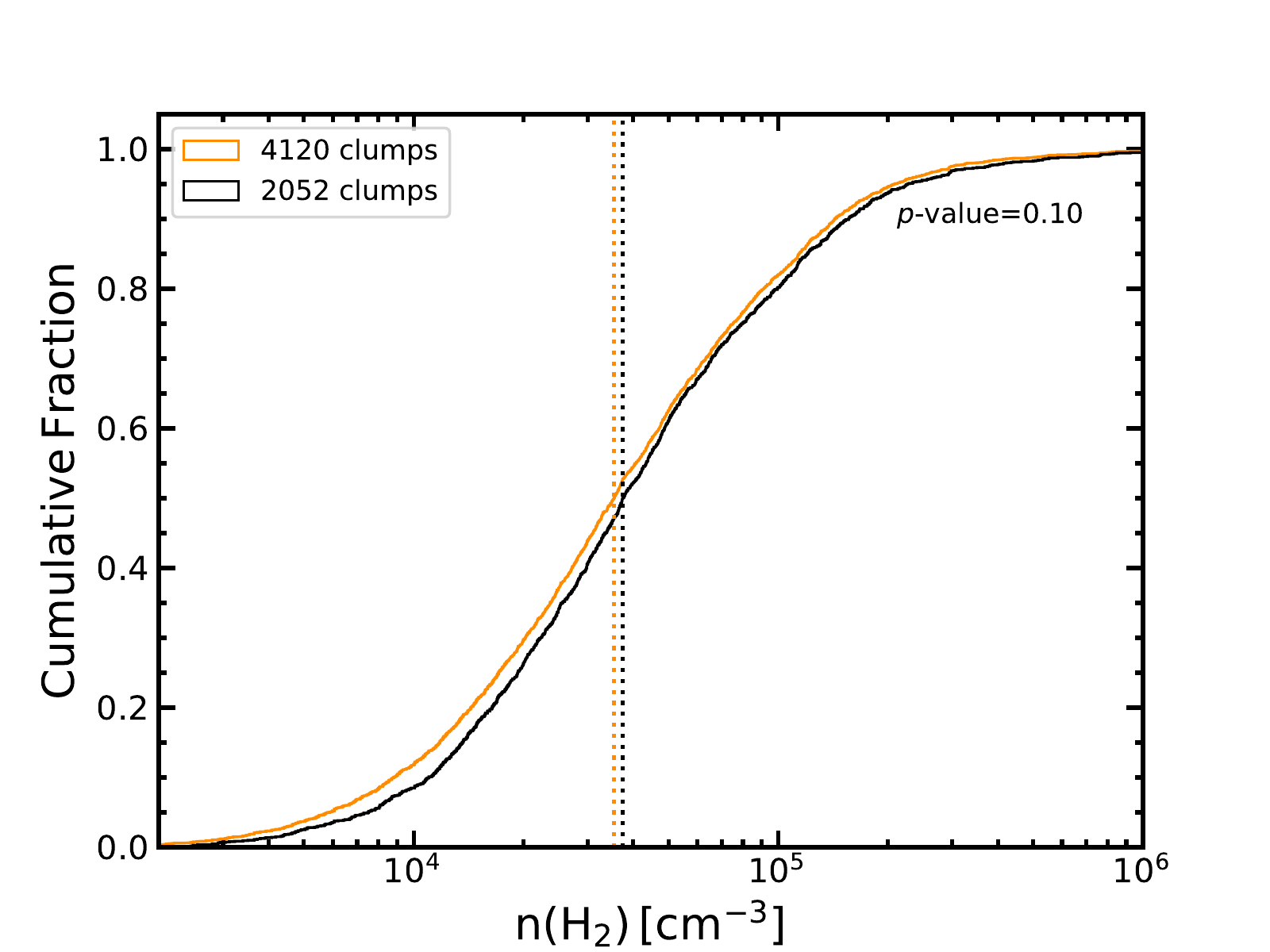} \\
 (d) & (e) & (f)   \\
\end{tabular}
 \caption{ Cumulative distributions of the physical properties for the total 4120 clumps and the selected 2052 clumps. Top left (a,b,c) to bottom right (d,e,f): Cumulative distributions of clump mass $\rm M_{clump}\,[M_{\odot}]$, bolometric luminosity of central objects $\rm L_{bol}\,[L_{\odot}]$, heliocentric distance\,Dist.\,[kpc], the peak $\rm H_2$ column density $\rm [peak\,N({H_2})/cm^{2}$], the mean mass surface density $\rm \Sigma\,[M_{\odot}/pc^{2}]$, and the mean volume $\rm H_2$ density $\rm n({H_2})\,[cm^{-3}]$ of the total 4120 clumps (orange  lines) and the selected 2052 clumps (black lines).  The orange and black vertical lines show the median values of the two samples for each parameter.
 }
\label{fig:cdf_distribution_param}
\end{figure*}

\begin{figure}
 \centering
 \includegraphics[width = 0.42\textwidth]{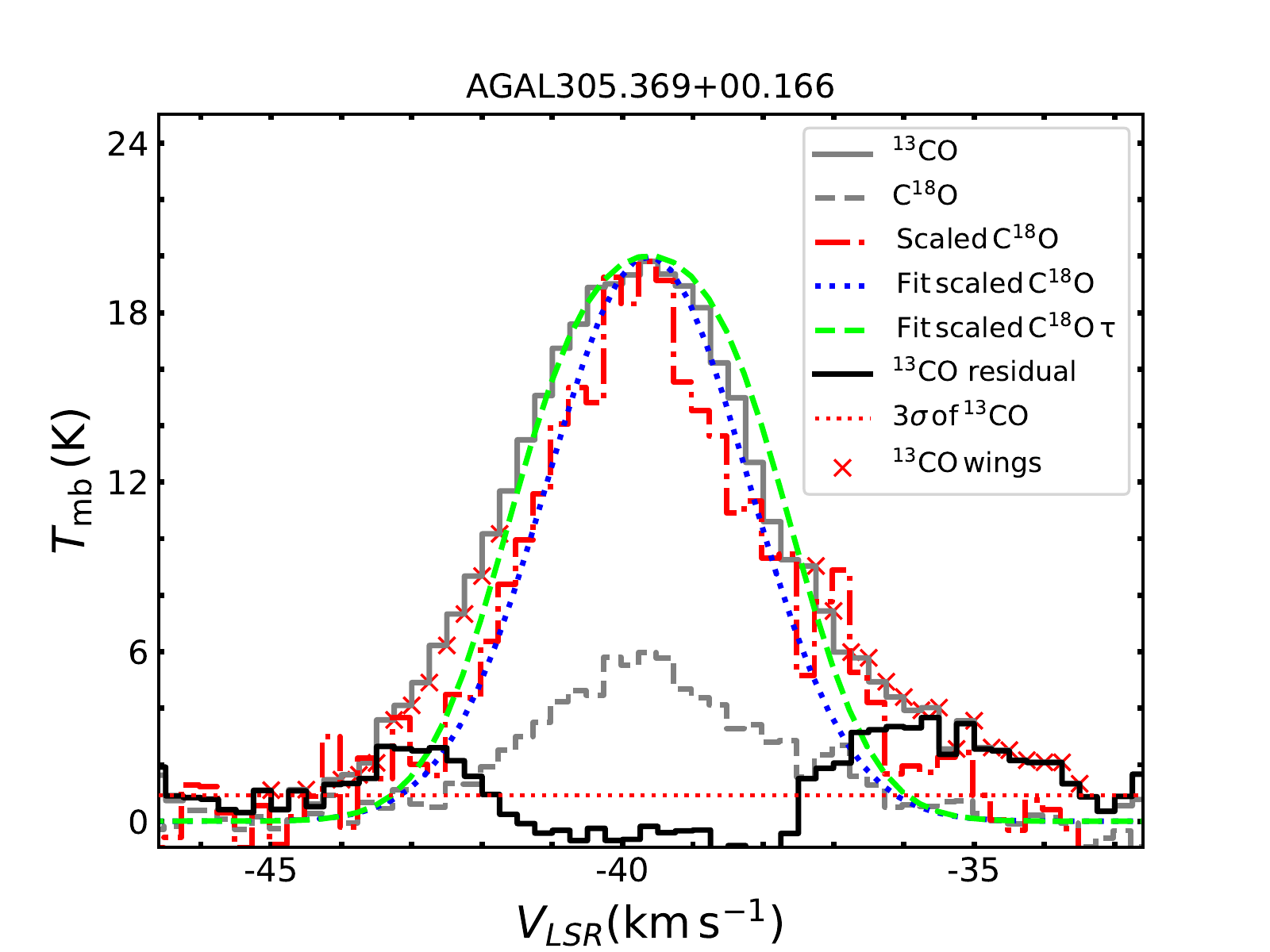}  \\
\includegraphics[width = 0.42\textwidth]{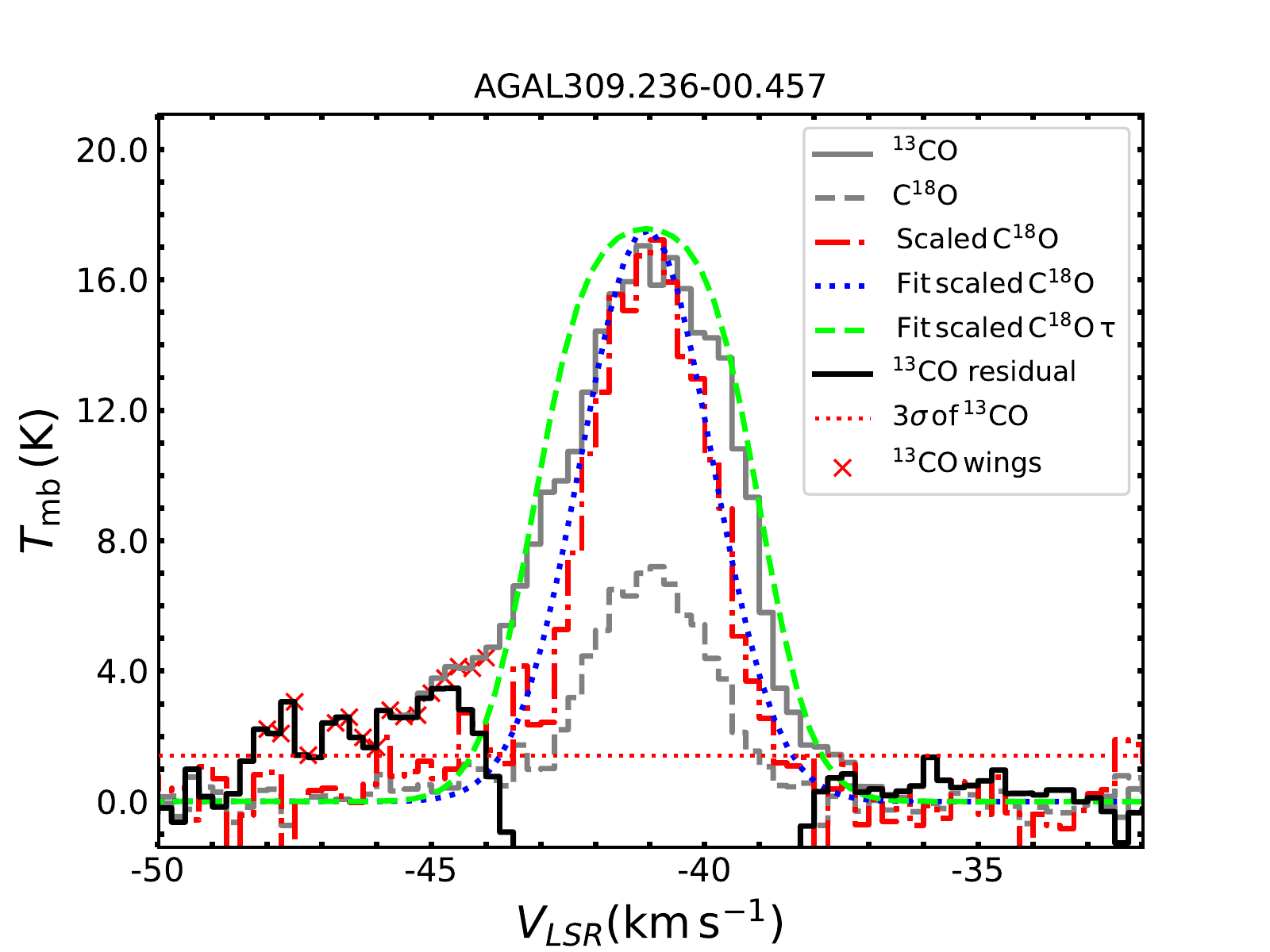}   \\
 \includegraphics[width = 0.42\textwidth]{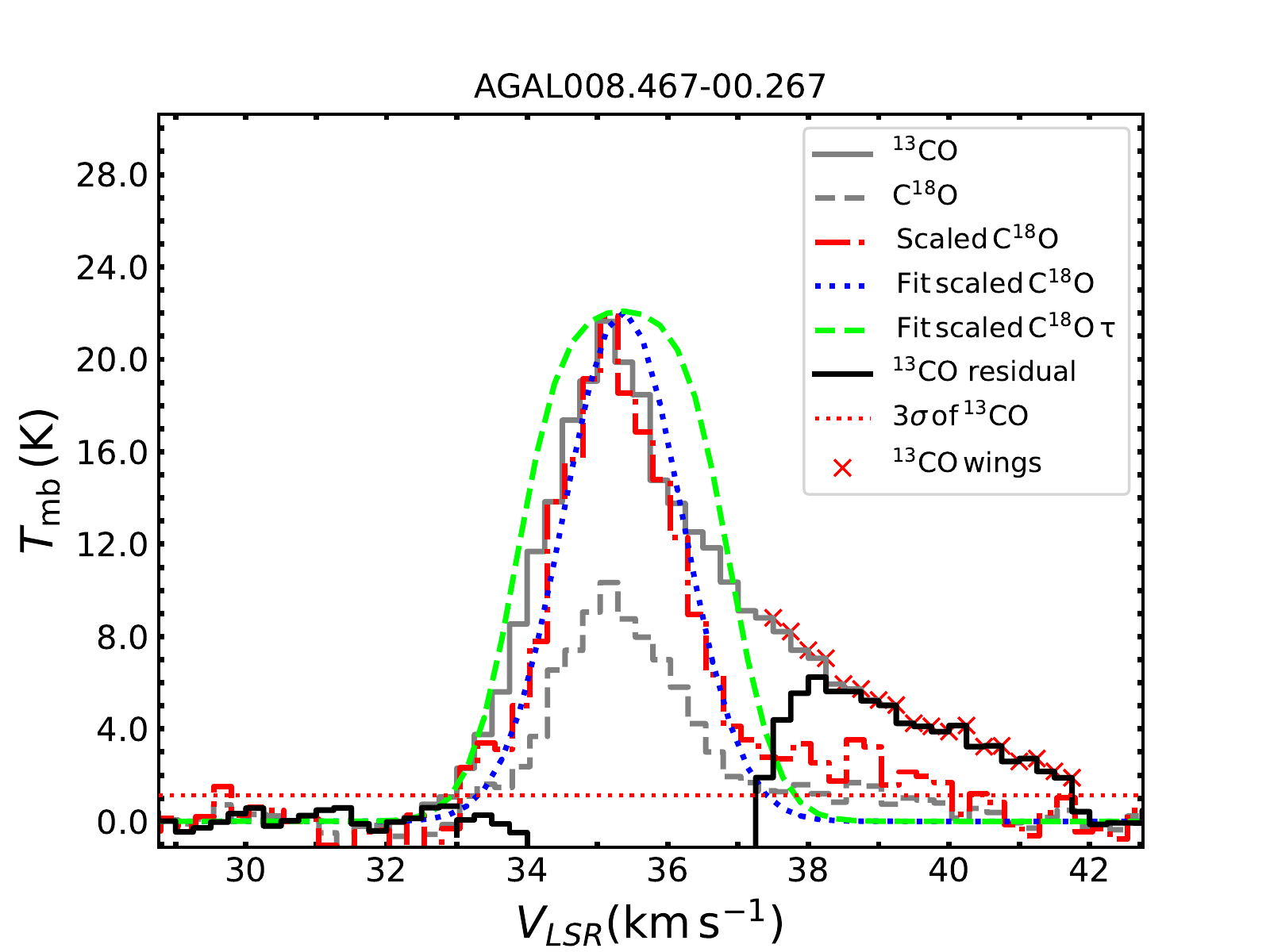}   \\
 \caption{Examples of outflow bipolar wings (top panel), unipolar blue wing (middle panel), and unipolar red wing (bottom panel). 
 The wings are selected by using spectra of the $\rm ^{13}CO$ (gray solid line) and $\rm C^{18}O$ (gray dashed line) for the ATLASGAL clump AGAL305.369$+$00.166, AGAL309.236-00.457, and AGAL008.467-00.267.
 Blue wings and red wings, shown as red cross symbols, can be identified following the procedures a b c, and d, in Sect.\,\ref{sect:wings_identification}. 
 }
\label{fig:wingselection}
\end{figure}
\section{The surveys and the clumps sample}
\label{sect:surveys}

\subsection{SEDIGISM}

The SEDIGISM\footnote{https://sedigism.mpifr-bonn.mpg.de/index.html \label{first_footnote}} survey covers a region of $-$60\degr\, $\leq$ $\ell$ $\leq$ +18\degr\ and $\vert b \vert$ $\leq$ 0.5\degr \citep{Schuller2017AA601A124S,Schuller2021MNRAS5003064S,Urquhart2021MNRAS3050U,Duarte_Cabral2021MNRAS3027D,Colombo2021arXiv211006071C} and was conducted using the Atacama Pathfinder Experiment (APEX) 12\,m submillimetre telescope \citep{Gusten2006AA454L13G}. 
The $^{13}$CO/C$^{18}$O ($J=2\rightarrow 1$) data used here are part of the Data Release 1 (DR1) data set (for details, see \citealt{Schuller2021MNRAS5003064S}), which has a typical $1\sigma$ rms noise of $\sim0.8$\,K (in $T_{\rm{mb}}$) per 0.25\,\kms\, channel and a full width at half maximum (FWHM) beam size of 28\arcsec. 
This rms allows for a $3\sigma$ detection that corresponds to column densities of $N_{\mathrm{H}_{2}} \sim 3 \times 10^{21}\,$cm$^{-2}$ (or 60\,\msunpc) for $^{13}$CO and $N_{\mathrm{H}_{2}} \sim 10^{22}\,$cm$^{-2}$ (or 200\,\msunpc) for C$^{18}$O, which is well suited to detect all molecular structures associated with star formation and their surrounding medium \citep{Schuller2017AA601A124S}. 
SEDIGISM is, therefore, a good tracer of the molecular gas associated with star-forming regions, with the $^{13}$CO line used to trace high-velocity structures.
The simultaneously observed C$^{18}$O line is typically optically thin compared to the $^{13}$CO in the same clump, and thus C$^{18}$O is an excellent tracer of emission from the dense cores  within clumps \citep[e.g.,][]{deVilliers2014MNRAS444,Yang2018ApJS235Y}.

\subsection{ATLASGAL} 

ATLASGAL is an unbiased 870\,$\rm \mu m$ submillimeter survey, covering $|\ell| \leq 60$ and $\vert b \vert$ $\leq$ 1.5\degr. 
ATLASGAL has a resolution of 19\arcsec\ and a typical noise level of 50$-$70\,mJy\,beam$^{-1}$ \citep{Schuller2009AA504, beuther2012, Csengeri2014AA565A75C}. 
A comprehensive database of $\sim 10163$ massive star-forming clumps has been compiled, the ATLASGAL compact source catalogue \citep[ CSC;][]{Contreras2013AA45C,Csengeri2014AA565A75C,urquhart2014c}, which allows us to undertake a search for CO outflow activity toward the clumps. 
Furthermore, the physical properties (e.g.,  distance, clump mass, column density, bolometric luminosity, density) have been measured and the evolutionary stages classified \citep{koenig2017AA599A139K,Urquhart2018MNRAS4731059U,Urquhart2021arXiv211112816U}. 
We use this well-characterized sample of clumps to conduct a statistical analysis of correlations between outflow parameters and clump properties for a large and representative sample of massive star-forming clumps in different evolutionary stages.

\subsection{The sample}
\label{sect:sample}
We used the common area of sky between the two surveys (i.e., $-$60\degr\, $\leq$ $\ell$ $\leq$ +18\degr\ and $\vert b \vert$ $\leq$ 0.5\degr) to conduct a search for outflows. 
We excluded the Galactic Center region\,($\vert \ell \vert < 5\degr$) due to the complexities of the gas kinematics in this part of the Galaxy \citep{Schuller2021MNRAS5003064S}. 
This provides a search area of 68\,deg$^{2}$ within which there are a total of 4120 ATLASGAL clumps. 

We extracted the \coa\ and \cob\ spectra from the SEDIGISM data for all of these clumps using the peak position and an area of the source size of each clump. 
Following the work presented in 
Paper\,I, 
we require that the emission is detected in both \coaa\ and \cobb . 
We found a final sample of 2052 clumps that fulfilled these criteria, as listed in Table\,\ref{tab:clump_properties}; this corresponds to approximately 50\% of the ATLASGAL sources in the overlapping region. 
We excluded 99 clumps (i.e., $\sim$2\% of the total sample) as they show complex $^{13}$CO spectra with multiple peaks (i.e., peaks > 2). 
Therefore, a final sample of 2052 clumps is obtained.
As shown in Fig.\,\ref{fig:galactic_distribution}, we overlaid the surveyed region (white shaded), the coordinates of the total 4120 clumps (black dots), and the selected 2052 clumps (yellow circles) on the artist's impression image of the Galactic structure to put the sampled clumps into a Galactic context. 
The physical properties of these clumps are presented in Table\,\ref{tab:clump_properties}, which were calculated in \citet{Urquhart2018MNRAS4731059U} and \citet{Urquhart2021arXiv211112816U}.
For the properties calculated in both \citet{Urquhart2018MNRAS4731059U} and \citet{Urquhart2021arXiv211112816U}, we adopted the values in \citet{Urquhart2021arXiv211112816U}, as they were recalculated using the FWHM sizes (determined from the pixels above the half-power level), to eliminate any observational bias that would make the clumps appear to increase in size and have decreasing volume densities with evolution. Full details are available in Fig. 33 of Urquhart et al. (2018) and Sect. 4.2 of \citet{billington2019_meth}.

From the selection criteria that require the detection of both \coaa\ and \cobb\ as described above, we are likely to select near and bright clumps in the total sample. 
This can be seen from the cumulative distributions of distance (Dist.) between the two samples in Fig.\,\ref{fig:cdf_distribution_param}, showing that the selected clumps are located systematically closer than the total clump sample. 
Also, the selected clumps have systematically higher values of bolometric luminosity [$\rm L_{bol}/L_{\odot}$] and peak $\rm H_{2}$ column density (peak $\rm N(H_{2})\,[cm^{2}$], derived from the peak flux density), compared to the total sample, as shown in Fig.\,\ref{fig:cdf_distribution_param}. 
Kolmogorov$-$Smirnov (K-S) tests between the two samples ($p$-value$\ll$0.001) also suggest that they are significantly different in heliocentric distance, bolometric luminosity, and peak $\rm N(H_{2}) [cm^{2}$]. 
However, the selected sample shows no significant differences from the total sample for the mean mass surface density ($\rm \Sigma\,[M_{\odot}/pc^{2}]$) and mean volume density ($\rm n(H_{2})/cm^{3}$) from the K-S tests and the cumulative distributions presented in Fig.\,\ref{fig:cdf_distribution_param}. 

In Table\,\ref{tab:summary_param} we provide a statistical summary of the physical properties for the full sample of clumps in the target region and the selected sample that passes the selection criteria of the detection of both \coaa\ and \cobb\ . Comparing these properties reveals that the selected clumps have comparable minimum, maximum, mean and median values of clump mass ($\rm M_{clump}/M_{\odot}$), bolometric luminosity ($\rm L_{bol}/L_{\odot}$), peak $\rm H_{2}$ column density [peak $\rm N(H_{2})/cm^{2}$], mean volume density ($\rm n(H_{2})/cm^{3}$) and mean mass surface density [$\rm\Sigma\,(M_{\odot}\,pc^{-2})$], compared to the full clump sample. 
Therefore, despite the inevitable selection bias due to the detection limit, the selected clump sample used to search for outflows, in general, has similar properties as the parent sample and may be taken to be representative of the total clump sample.

\section{Outflow wing identification} 
\label{sect:wings_identification}

The strategy of identifying outflow wings builds on the method from Paper\,I and \citet{deVilliers2014MNRAS444}, which relies on using optically thin \cobb\ to trace the dense gas and \coaa\ to detect the high-velocity gas. 
In this work, we extract \coa\ and \cob\ spectra from the SEDIGISM data cubes, integrated over the area of each clump, centered on the peak emission of each ATLASGAL source.
The high-velocity outflow wings are defined by the velocities where the observed $\rm ^{13}CO$ profile is broader than the scaled \cobb\ line  (as discussed below) representing core emission \citep[e.g.,][]{Codella2004AA615C,vanderWalt2007AA464,deVilliers2014MNRAS444,Yang2018ApJS235Y}. 
Observed line profiles can be affected by line opacity \citep{Hacar2016AA591A104H}, with these effects previously reported in studies of the J=2$-$1 transition in \coaa\ \citep[e.g.,][]{Wilson1999ApJ174W}. Therefore, the broad wings derived from the \coa\ line may be caused by this opacity broadening. As a result, we developed a method of identifying outflow wings in this work aiming to properly subtract the corresponding opacity broadening from the observed line. 
This is achieved by appending a new step before the subtraction procedure of the original method in Paper\,I, in which the line broadening effects are added to the line representing core-only emission. 

We illustrate the basic procedures of the developed method in Fig.\,\ref{fig:wingselection}. 
As mentioned above, the \coaa\ and \coab\ spectra for all clumps are extracted from the SEDIGISM data at their peak positions averaged over the area of clump sizes reported by \citet{Urquhart2018MNRAS4731059U}. 
Starting from the observed spectra of \coaa\ (gray solid line) and \cobb\ (gray dashed line) of the ATLASGAL clumps, the steps to identify outflow wings are (a) scaling the \cobb\ line to the peak temperature of the \coaa\ line (red dash-dotted line); (b) fitting a Gaussian to the scaled \cobb\ line (blue dotted line); (c) adding opacity broadening effects to (b), the Gaussian fit to the scaled \cobb\ line (lime dashed line); (d) obtaining \coaa\ residuals (in black solid line), by subtracting the scaled Gaussian $\rm{C^{18}O}$ line with opacity broadening (lime dashed line) from the observed $\rm{^{13}CO}$ line (gray solid line); (e) identifying the blue and red line wings (red cross symbols) where the $^{13}$CO residuals are larger than  3$\sigma$, where $\sigma$ is the noise level of the emission-free regions of the corresponding spectrum for each clump. 

The added step (c) is the process to account for the contribution from the opacity broadening to the Gaussian fit to the scaled \cobb\ line (core-only emission) of step (b). 
This can be achieved by applying the central optical depth of \coa\ to the equations (1) and (2) in \citet{Hacar2016AA591A104H}. 
To be specific, the line opacity is assumed to have a Gaussian distribution in frequency ($\nu$) and velocity ($V_{lsr}$), that is, the equation (2) in \citet{Hacar2016AA591A104H}, 
$\rm \tau_{\nu}=\tau_0 \cdot exp(-(\nu-\nu_0)^2/2\sigma^2)$ and $\frac{\nu-\nu_0}{\nu_0} = \frac{V_{lsr,0}-V_{lsr}}{c}$. 
Where $\tau_0$, $\nu_0$, and $V_{lsr,0}$ are the central opacity, line frequency, and line-of-sight velocity, respectively, and $\sigma$ is the intrinsic velocity dispersion of the gas, namely, the FWHM of the scaled Gaussian \cobb\ line. 
Then, applying $\tau_{\nu}$, the emission distribution of the molecular line \citep{Wilson2013trabookW}, that is, equation (1) in \citet{Hacar2016AA591A104H}, is $\rm T_{mb,\nu}=(J_\nu(T_{ex})-J_\nu(T_{bg}))\cdot(1-\exp(-\tau_\nu))$, with radiation temperature $J_\nu(T)=\left(\frac{h\nu/k}{exp(h\nu/kT)-1}\right)$, T$_{ex}$ being the excitation temperature of
the line, T$_{bg}$ the cosmic microwave background temperature
(=2.7 K). Here, the $J_\nu(T_{ex})-J_\nu(T_{bg})$ refers to the intensity of the Gaussian fit to the scaled \cobb\ (blue-dotted line in Fig\,\ref{fig:wingselection}), as mentioned above.
For each source, an estimate of the central optical depth $\tau_{0}$ of \coa\ has been obtained from the ratio of the intensities of \coa\ and \cob, assuming that \cob\ is optically thin and a CO abundance ratio X(\coaa)/X(\cobb)=7.3 \citep{Goldsmith1984ApJ286599G,Wilson1994ARAA191W,Jim2017MNRAS46649J}. 
The central optical depth $\tau_{0}$ of the sample has a median value of 2.8, ranging from 0.3 to 7.4, indicating that the contributions of opacity broadening to the line-widths of \coaa\ are expected to be from $\sim$5\% to $\sim$50\% (see figure 2 of \citet{Hacar2016AA591A104H}). 
Step (d) as outlined above is to subtract both the opacity broadening and core emission (i.e., the lime line in Fig.\,\ref{fig:wingselection}) from the observed \coaa\ profile. This removes the \coaa\ residuals above 3$\sigma$ from step (d),  allowing the observed broad wings to be dominated by the kinematics of the gas rather than the corresponding line opacity. 
Following the above procedures, blue wings ($-$45.0\kms\ to $-$41.75\kms) and red wings ($-$37.25\kms\ to $-$33.5\kms) of outflow activity are determined for AGAL305.369$+$00.166, as displayed in Fig.\,\ref{fig:wingselection}. 
 



\begin{table*}
\centering
\caption {\rm Summary of physical parameters of clumps and outflows. In Columns (2-5) we give the minimum, maximum, $\rm mean\pm standard\,deviation$, and median values of these  parameters for each subsample. 
}
\begin{tabular}{lllll}
\hline
\hline
Parameter  &  $x_{min}$ & $x_{max}$ & $x_{mean}\pm x_{std}$ & $x_{med}$\\
\hline
\multicolumn{5}{c}{4120 ATLASGAL clumps in SEDIGISM }     \\
\hline
$\rm \log(M_{clump}/M_{\odot})$ & -1.52 & 4.39 & $2.52\pm0.62$& 2.53 \\
$\rm \log(L_{bol}/L_{\odot})$ & -0.63 & 6.29 & $2.94\pm1.03$ & 2.89 \\
$\rm \log[L_{bol}/M_{clump}(L_{\odot}/M_{\odot})]$ & -2.0 & 3.25 & $0.42\pm0.88$ & 0.40 \\
$\rm \log(Peak\,N({H_2})/cm^{2})$ & 21.58 & 23.80 & $22.31\pm0.29$  & 22.28 \\
$\rm \log[mass\,surface\,density\,\Sigma(M_{\odot}\,pc^{-2})]$ & 2.16 & 4.79 & $3.04\pm0.29$ & 3.02 \\
$\rm \log(n(H_{2})/cm^{-3})$ & 2.95 & 6.55 & $4.56\pm0.48$ & 4.55 \\

\hline
\multicolumn{5}{c}{2052 clumps with detection in \coaa\ and \cobb\ }     \\
\hline
$\rm \log(M_{clump}/M_{\odot})$ & -1.22 & 4.39 & $2.49\pm0.56$& 2.48 \\
$\rm \log(L_{bol}/L_{\odot})$ & 0.0 & 6.29 & $3.18\pm1.03$ & 3.14 \\
$\rm \log[L_{bol}/M_{clump}(L_{\odot}/M_{\odot})]$ & -1.70 & 2.94 & $0.69\pm0.84$ & 0.70 \\
$\rm \log(Peak\,N({H_2})/cm^{2})$ & 21.58 & 23.80 & $22.38\pm0.32$  & 22.34 \\
$\rm \log[mass\,surface\,density\,\Sigma(M_{\odot}\,pc^{-2}]$ & 2.23 & 4.79 & $3.06\pm0.30$ & 3.03 \\
$\rm \log(n(H_{2})/cm^{-3})$ & 3.27 & 6.55 & $4.60\pm0.47$ & 4.57 \\
\hline
\multicolumn{5}{c}{ 1192 clumps with outflows}     \\
\hline
$\rm \log(M_{clump}/M_{\odot})$ & -1.22 & 4.39 & $2.53\pm0.54$ & 2.52 \\
$\rm log(L_{bol}/L_{\odot})$ & 0.0 & 6.08  & $3.34\pm1.04$ & 3.30 \\
$\rm \log[L_{bol}/M_{clump}(L_{\odot}/M_{\odot})]$ & -1.70 & 2.94 & $0.82\pm0.85$ & 0.84 \\
$\rm \log(Peak\,N({H_2})/cm^{2})$ & 21.58 & 23.80 & $22.43\pm0.34$  & 22.38 \\
$\rm \log[mass\,surface\,density\,\Sigma(M_{\odot}\,pc^{-2}]$ & 2.23 & 4.79 & $3.09\pm0.32$ & 3.05 \\
$\rm \log(n(H_{2})/cm^{-3})$ & 3.40 & 6.55 & $4.63\pm0.47$ & 4.62 \\
\hline
\multicolumn{5}{c}{ 860 clumps without outflows}     \\
\hline
$\rm log(M_{clump}/M_{\odot})$ & 0.10 & 4.33 & $2.45\pm0.57$ & 2.44 \\
$\rm log(L_{bol}/L_{\odot})$ & 0.23 & 6.29 & $2.97\pm0.97$ & 2.91 \\
$\rm \log[L_{bol}/M_{clump}(L_{\odot}/M_{\odot})]$ & -1.70 & 2.79 & $0.52\pm0.79$ & 0.50 \\
$\rm \log(Peak\,N({H_2})/cm^{2})$ & 21.73 & 23.41 & $22.31\pm0.28$  & 22.27 \\
$\rm \log[mass\,surface\,density\,\Sigma(M_{\odot}\,pc^{-2})]$ & 2.36 & 4.00 & $3.01\pm0.27$ & 2.99 \\
$\rm \log(n(H_{2})/cm^{-3})$ & 3.27 & 5.97 & $4.55\pm0.46$ & 4.53 \\
\hline
\hline
\end{tabular}
\label{tab:summary_param}
\end{table*} 

\begin{table*}
\centering
\caption[]{\it \rm $\rm{^{13}CO}$ outflow calculations of all blue and red wings for 1192 ATLASGAL clumps: 
observed peak $\rm{^{13}CO}$ and $\rm C^{18}O$ velocities, main-beam temperatures, velocity intervals, $[V_{min_{b/r}},V_{max_{b/r}}]$ for blue and red wings of $\rm{^{13}CO}$ spectra, maximum projected velocity $\rm \Delta V_{max_{b/r}}$ for blue and red shifted relative to the peak $\rm C^{18}O$ velocity, that is, $\rm \Delta V_{max_{b}}=V_{c^{18}o}-V_{min_{b}}$ and $\rm \Delta V_{max_{r}}=V_{max_{r}}-V_{c^{18}o}$. Only a small part of the table is presented here, with a full version available from CDS.}
 
\begin{tabular}{lllllllll}
\hline
\hline
ATLASGAL            &$\rm{^{13}CO}\,v_p$ & $\rm{^{13}CO}\,T_{mb}$ & $\rm C^{18}O\,v_p$ & $\rm C^{18}O\,T_{mb}$ & $\rm[V_{min_{b}},V_{max_{b}}]$ &  $\rm[V_{min_{r}},V_{max_{r}}]$ &  $\rm \Delta V_{max_{b}}$ & $\rm \Delta V_{max_{r}}$ \\
CSC Gname          &   ($\rm km\,s^{-1}$)    & (K)                                      &   ($\rm km\,s^{-1}$)    & (K) &   ($\rm km\,s^{-1}$) &   ($\rm km\,s^{-1}$)  &   ($\rm km\,s^{-1}$)    &   ($\rm km\,s^{-1}$) \\ 
\hline
AGAL005.001$+$00.086 & 2.1 & 5.1 & 2.1 & 2.2 & [$-$0.25,1.0] & [3.5,7.25] & 2.4 & 5.1 \\
AGAL005.321$+$00.184 & 8.5 & 9.2 & 8.5 & 4.4 & [6.5,7.0] & [10.5,11.5] & 2.0 & 3.0 \\
AGAL005.329$+$00.181 & 8.6 & 9.8 & 8.8 & 4.1 & [6.5,7.5] & [10.25,12.0] & 2.3 & 3.2 \\
AGAL005.352$+$00.116 & 11.2 & 7.7 & 11.0 & 3.3 & [9.5,9.75] & $-$ & 1.5 & 0.0 \\
AGAL005.354$+$00.094 & 12.0 & 9.9 & 11.8 & 4.8 & [9.75,10.25] & $-$ & 2.1 & 0.0 \\
AGAL005.359$+$00.014 & 12.5 & 9.5 & 12.5 & 4.0 & [10.5,11.5] & [13.75,14.5] & 2.0 & 2.0 \\
AGAL005.387$+$00.187 & 11.2 & 7.3 & 11.2 & 3.8 & [9.5,10.0] & [12.5,13.0] & 1.7 & 1.8 \\
AGAL005.397$+$00.194 & 11.3 & 10.0 & 11.4 & 5.4 & [9.5,10.25] & [12.0,13.5] & 1.9 & 2.1 \\
AGAL005.637$+$00.237 & 8.5 & 13.6 & 9.2 & 6.7 & [3.5,6.75] & [11.75,17.0] & 5.7 & 7.8 \\
AGAL005.712$-$00.117 & 13.1 & 22.2 & 13.1 & 7.0 & [11.25,12.0] & [14.25,14.75] & 1.8 & 1.7 \\

  \hline
\hline
\end{tabular}
\label{tab:outflow_wings}
\end{table*}    

\begin{figure*}
\centering
\begin{tabular}{ccc}
 \includegraphics[width = 0.32\textwidth] {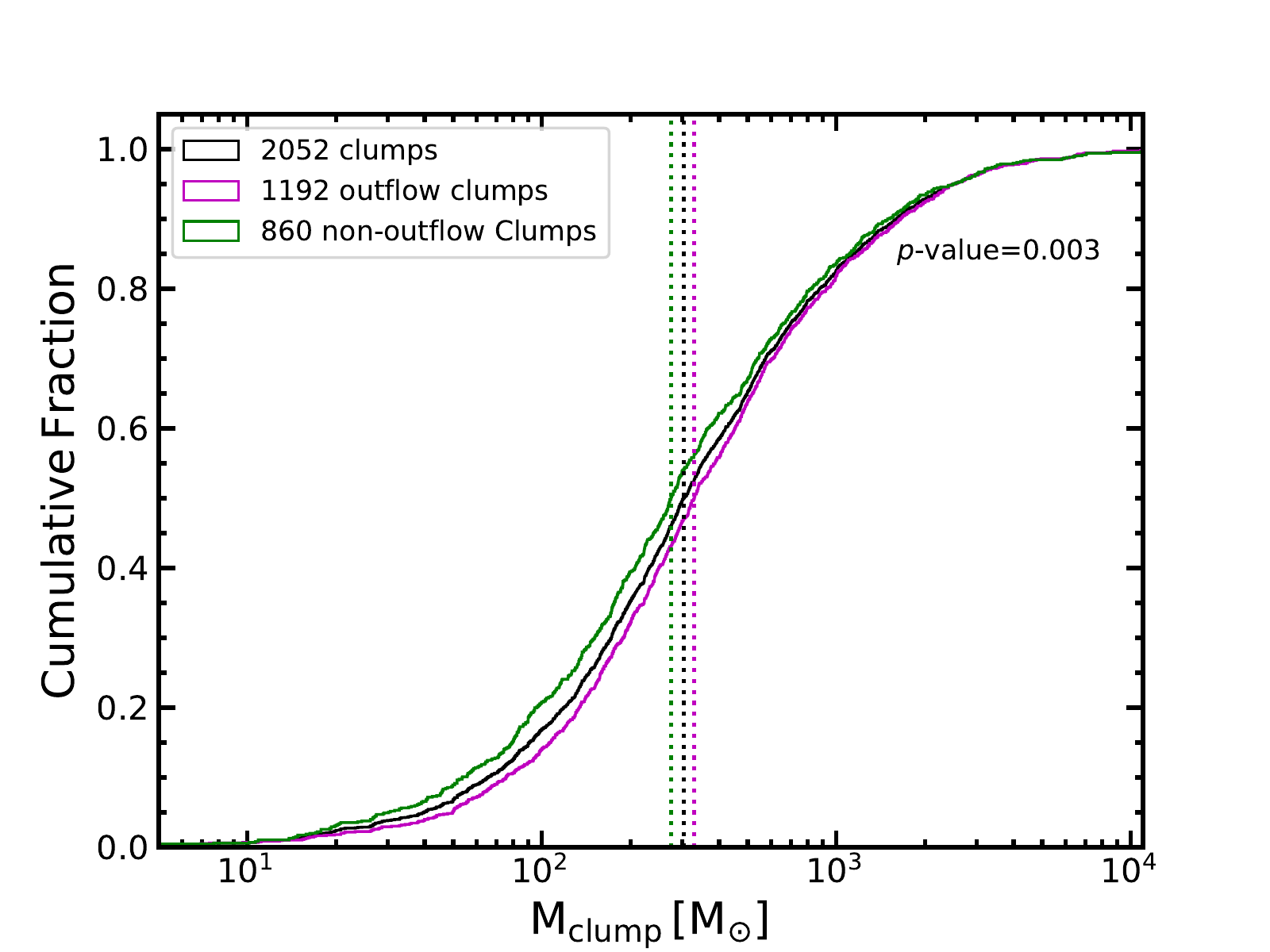} & 
 \includegraphics[width = 0.32\textwidth] 
 {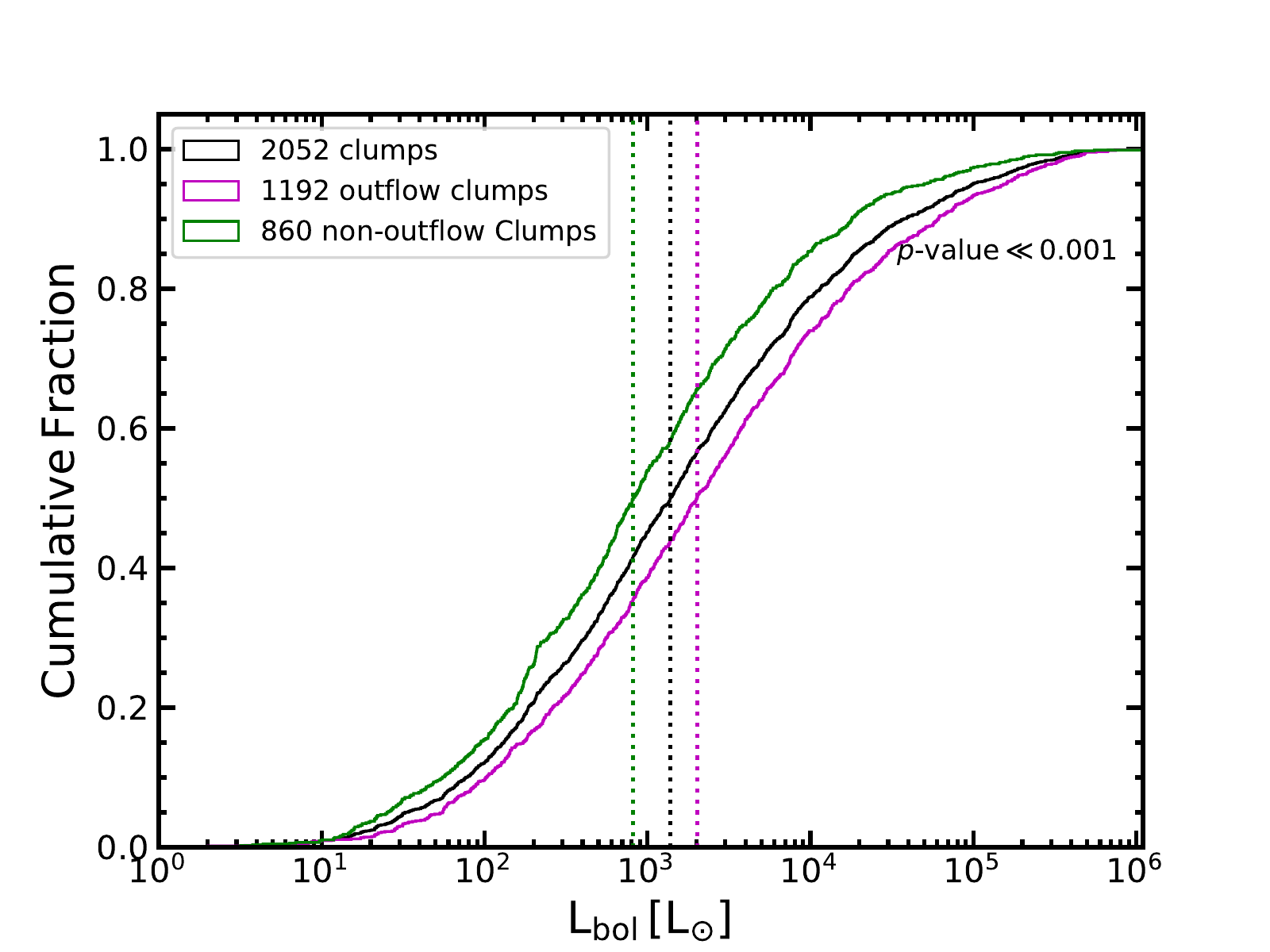} & 
 \includegraphics[width = 0.32\textwidth] 
 {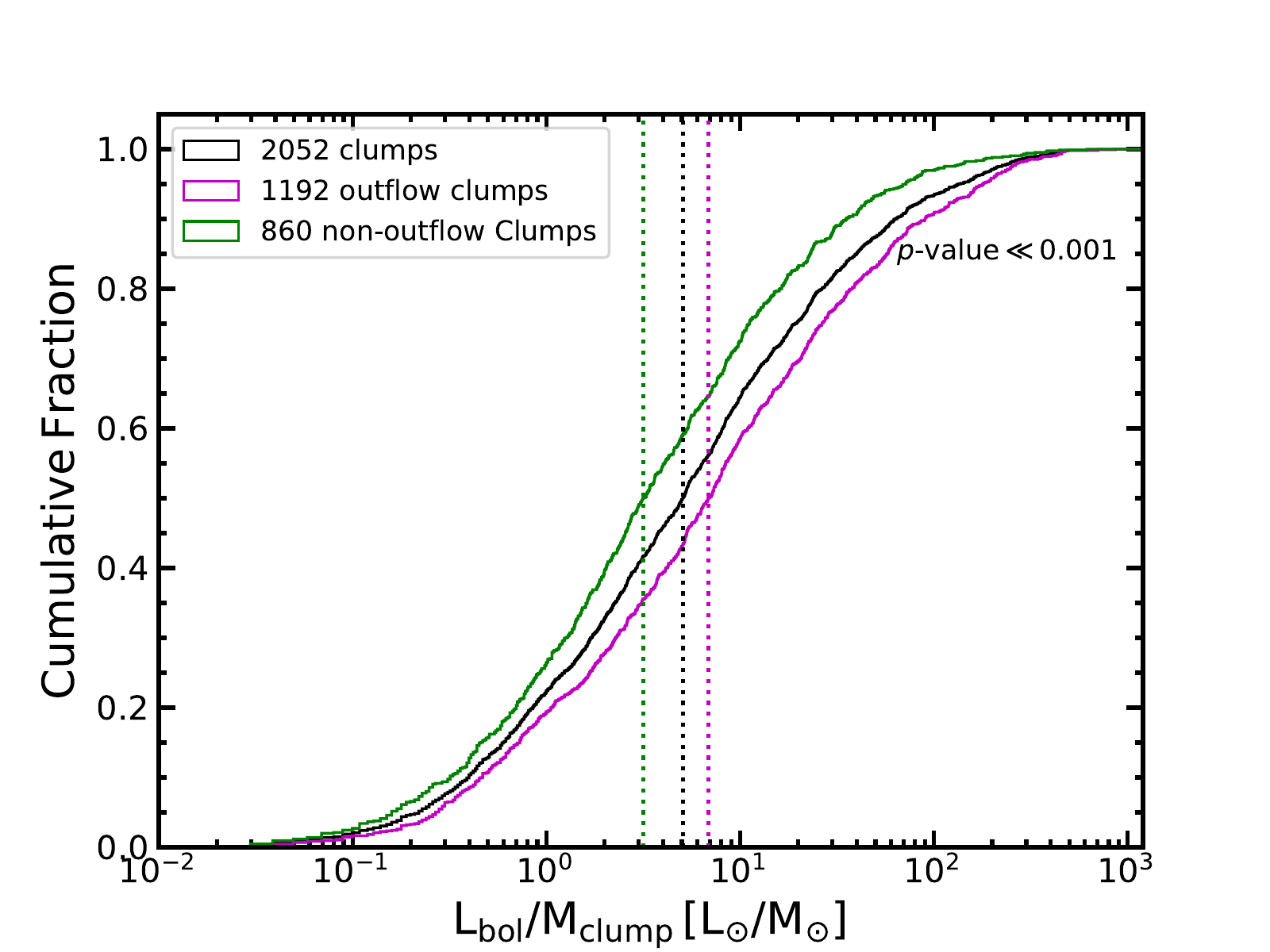} \\
 (a)& (b) & (c) \\
   \includegraphics[width = 0.32\textwidth] {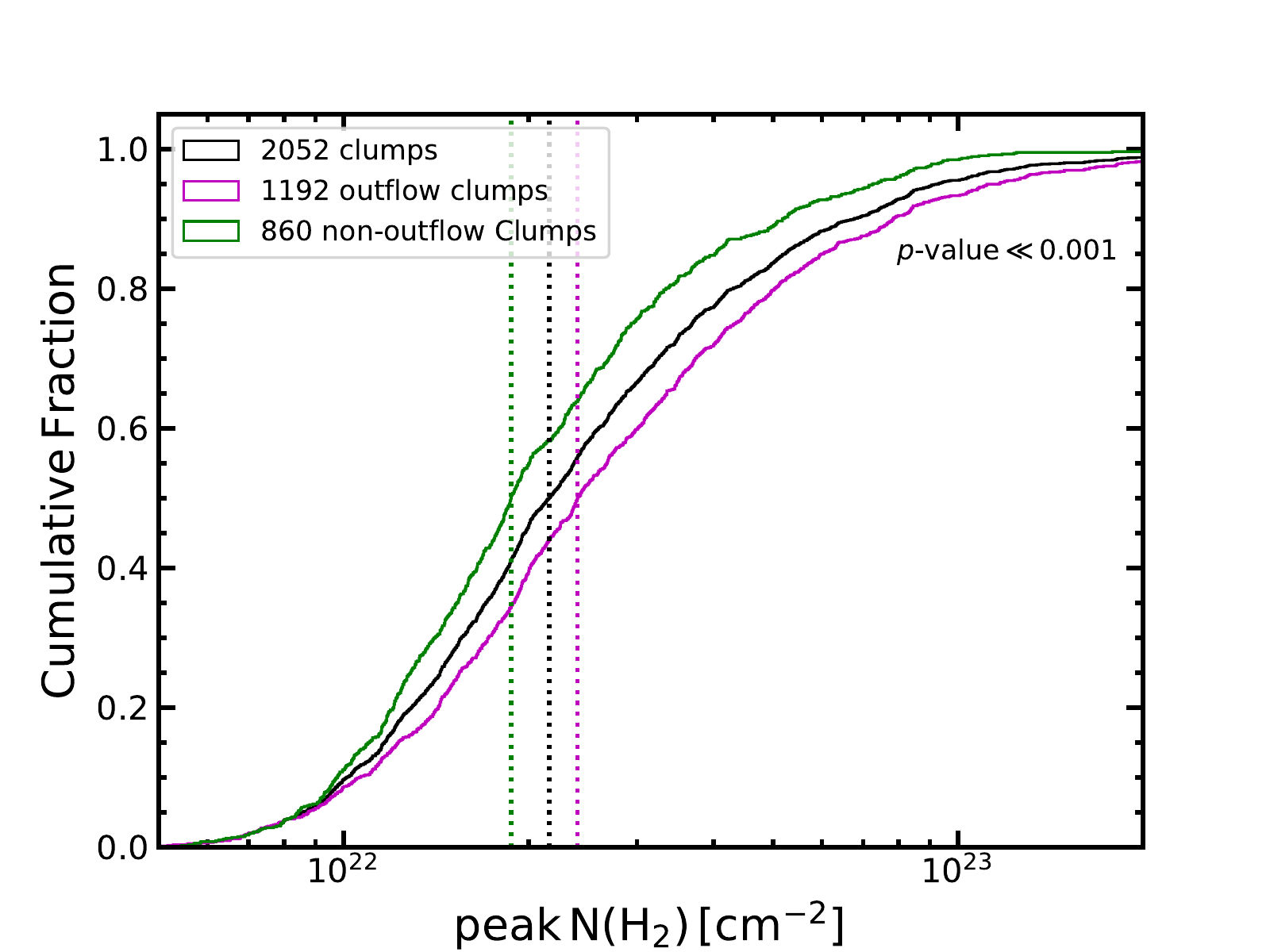} &
  \includegraphics[width = 0.32\textwidth] {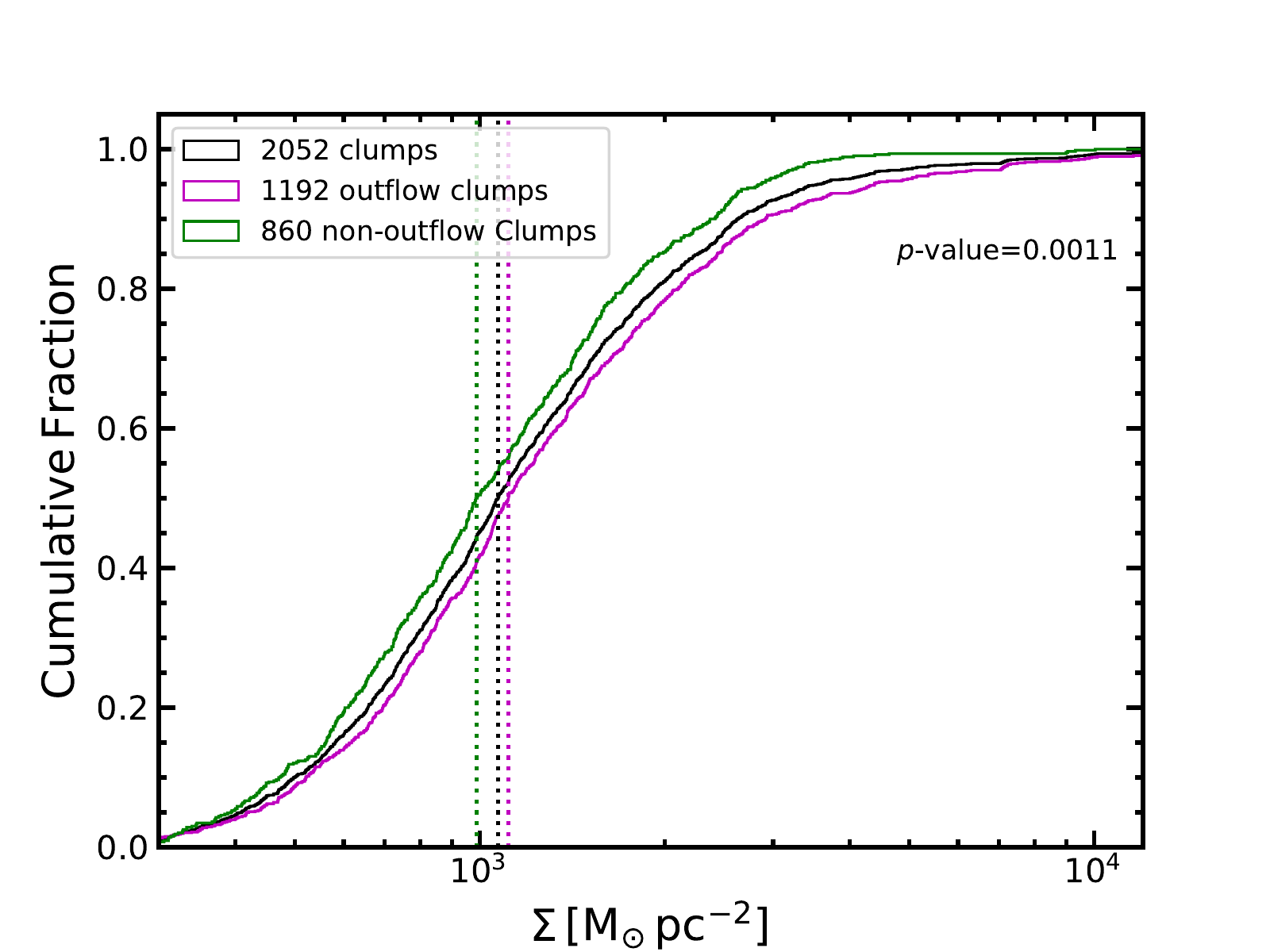} &
 \includegraphics[width = 0.32\textwidth] {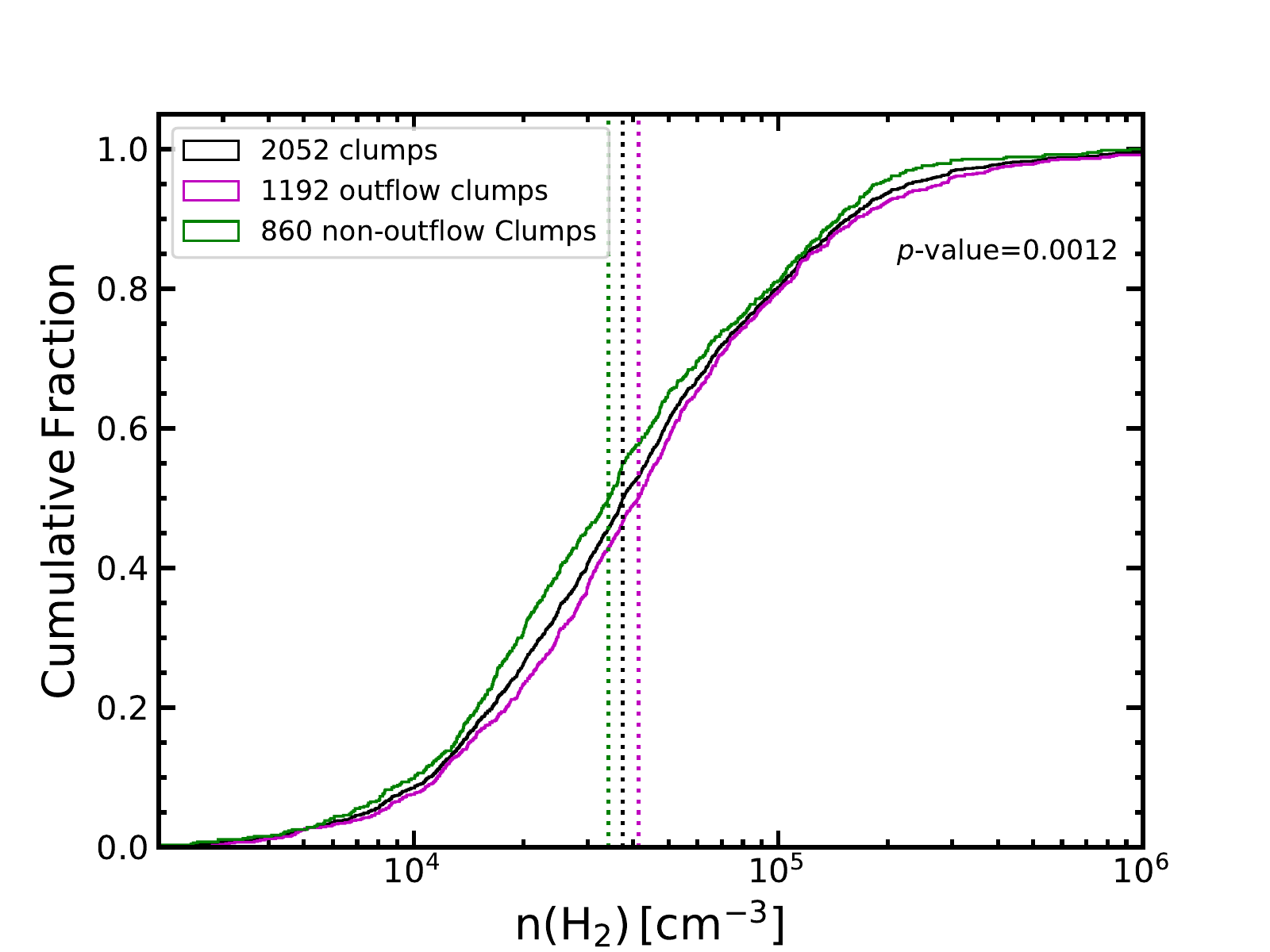} \\
 (d) & (e) & (f)  \\
\end{tabular}
 \caption{ Cumulative distributions of the physical properties for the total 2052 clumps, the 1192 outflow clumps, and the 860 non-outflow clumps. 
 Top left (a,b,c) to bottom right (d,e,f): Cumulative distributions of clump mass [$\rm M_{clump}\,(M_{\odot})$], bolometric luminosity of central objects [$\rm L_{bol}/L_{\odot}$], luminosity-to-mass ratio [$\rm L_{bol}/M_{clump}\,(L_{\odot}/M_{\odot})$], the peak $\rm H_2$ column density $\rm [peak\,N({H_2})/cm^{2}$], the mean mass surface density [$\rm \Sigma\,(M_{\odot}/pc^{2})$], and the mean volume $\rm H_2$ density [$\rm n({H_2})/(cm^{3})$] of the total 2052 clumps (black lines), the 1192 outflow clumps ( magenta lines), and the 860 non-outflow clumps (green lines). 
  We present the median values of the three samples in vertical lines, and $p$-values of K-S tests between the outflow sample and the non-outflow in each plot. 
 }
\label{fig:cdf_distribution_param_outflow}
\end{figure*}
\begin{figure*}
\centering
\begin{tabular}{cc}
 \includegraphics[width = 0.45\textwidth]{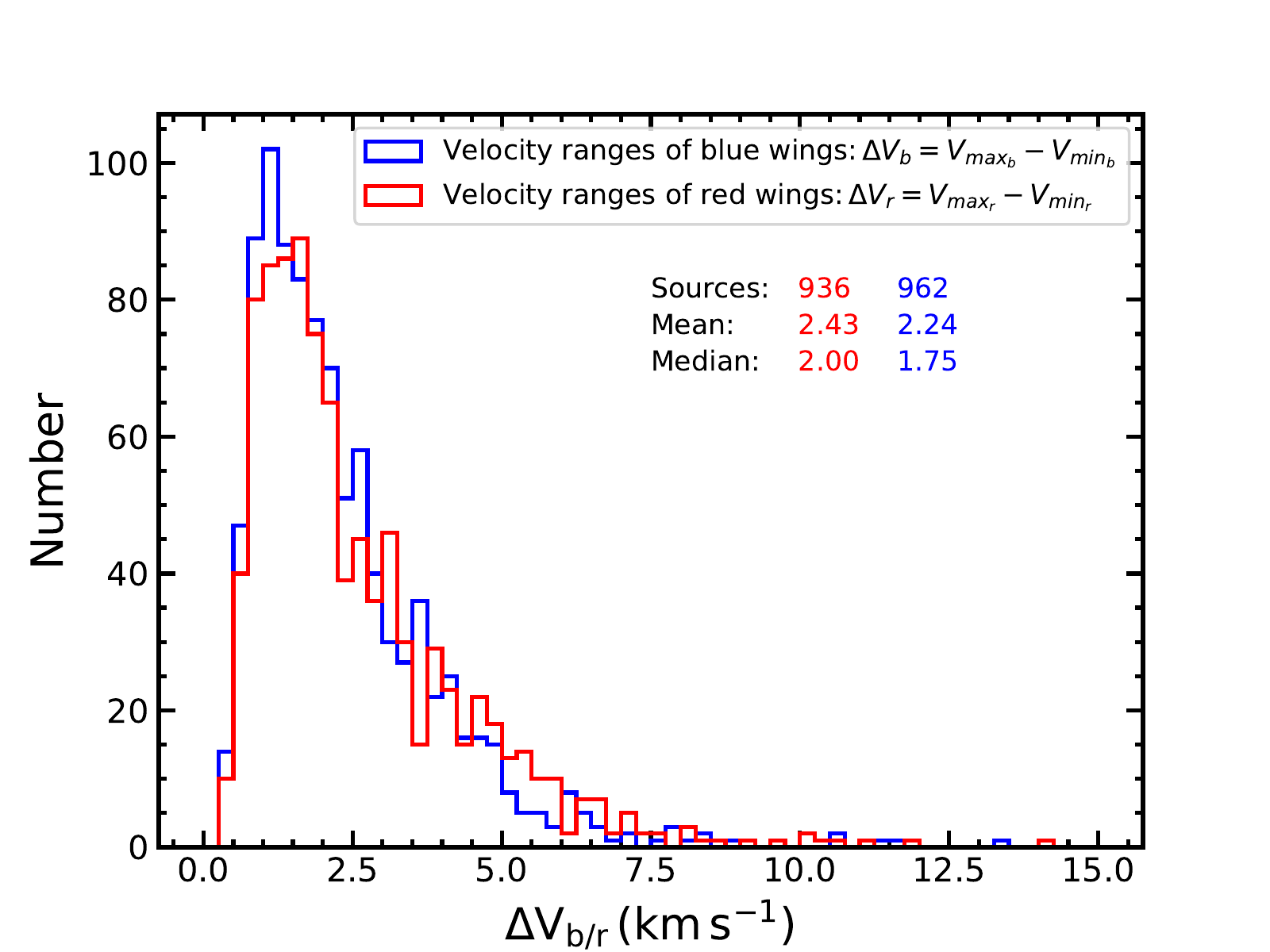} & 
 \includegraphics[width = 0.45\textwidth]{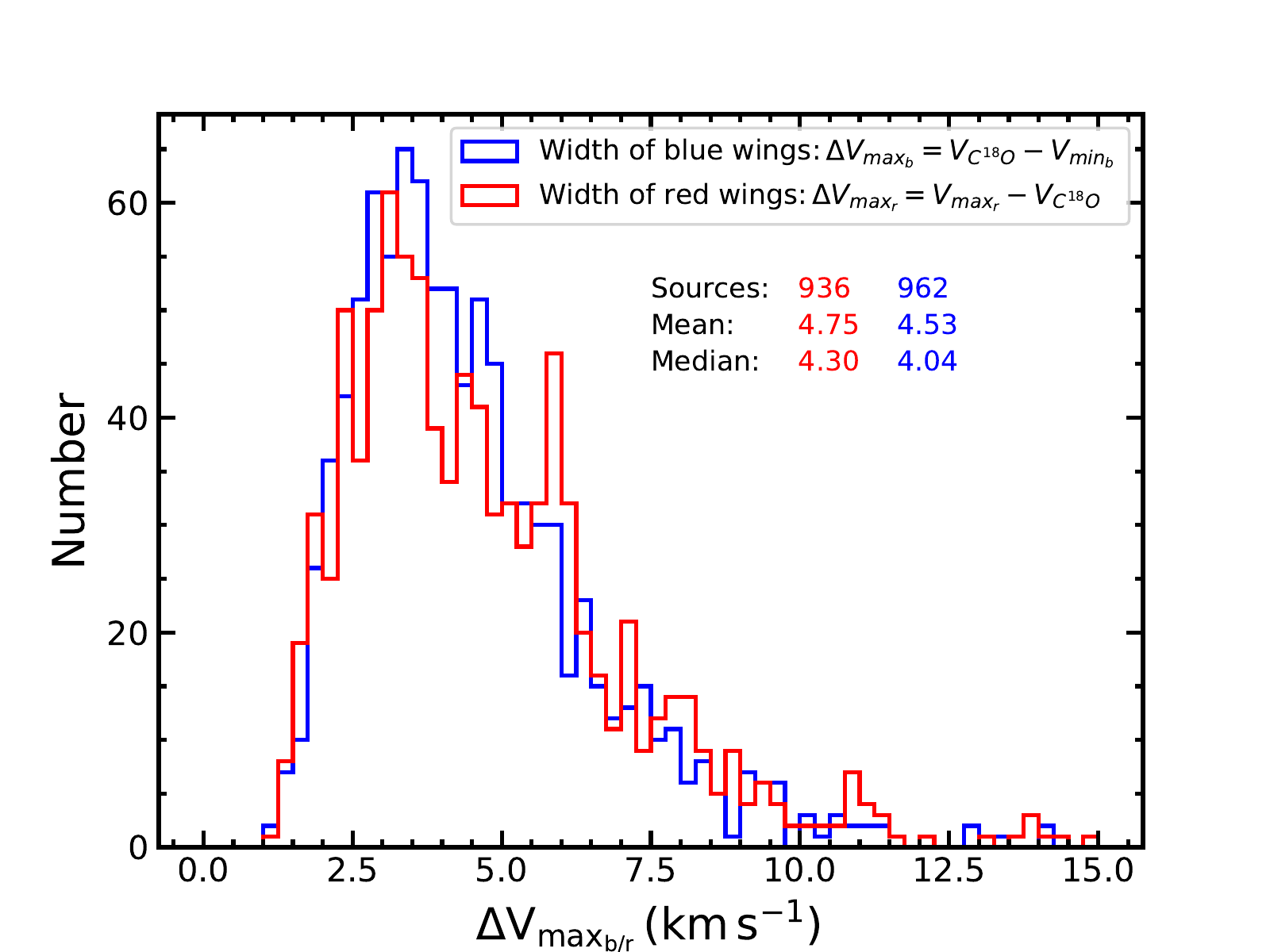} \\

\end{tabular}
 \caption{ Distributions of velocity ranges $\Delta V_{b/r}$ (left-panel), maximum blue and red wing velocities $\Delta V_{max_{b/r}}$ (right-panel) of the 1192 clumps with outflow wings, including 706 clumps with bipolar wings, and  486 clumps with  unipolar wings.
 }
\label{fig:wings_velocity}
\end{figure*}
\section{Results}

\label{sect:results}
Based on the search for outflows outlined above, 
we find that 1192 out of 2052 clumps are associated with high-velocity outflow signatures (i.e., the outflow candidates), and the remaining 860 clumps show no outflow wings (i.e., the non-outflow candidates).
It is likely that we missed some outflows as the outflow identification process can be affected by confusion (the observed sources lie along the Galactic plane where most of the molecular material resides), spectral noise (in the case of weak sources), and outflow geometry (which determines the width of the broad wings), as discussed in \citet{Codella2004AA615C} and Paper\,I. 
In addition, the beam filling factor (i.e., $\Omega_{source\,size}/\Omega_{beam\,size}$) of \coaa\ line emission for clumps at different distances \citep[e.g.,][]{Yan2021ApJ910109Y} and the opacity variations in the \coaa\ line wings \citep[e.g.,][]{Goldsmith1984ApJ286599G} can also affect the outflow identification process in this work. 
However, given the homogeneity of the present sample and the large number of observed clumps, the results should be representative of the general population and therefore give an accurate picture of the commonality of outflows and their properties.

A summary of the physical properties of the outflow and non-outflow sample are listed in Table\,\ref{tab:summary_param}. 
The 1192 outflow clumps have slightly higher mean and median values of mass and bolometric luminosity as well as peak column, mean mass surface, and mean volume densities of $\rm H_{2}$ compared to the 860 non-outflow sample, as displayed in Fig.\,\ref{fig:cdf_distribution_param_outflow}. K-S tests of these physical properties confirm that the two samples are significantly different from each other, with $p$-values $\ll$ 0.013, as shown in Fig.\,\ref{fig:cdf_distribution_param_outflow}.
Among the 1192 outflow clumps, 706 of them show bipolar wings, that is, associated with both red and blue high-velocity structures, and the remaining 486 clumps show unipolar red and blue high-velocity wings. 
No significant differences are found for the two subsamples: clumps with bipolar wings and clumps with unipolar wings, when K-S tests are performed.

\begin{figure}[!htp]
 \centering
   \includegraphics[width = 0.49\textwidth]{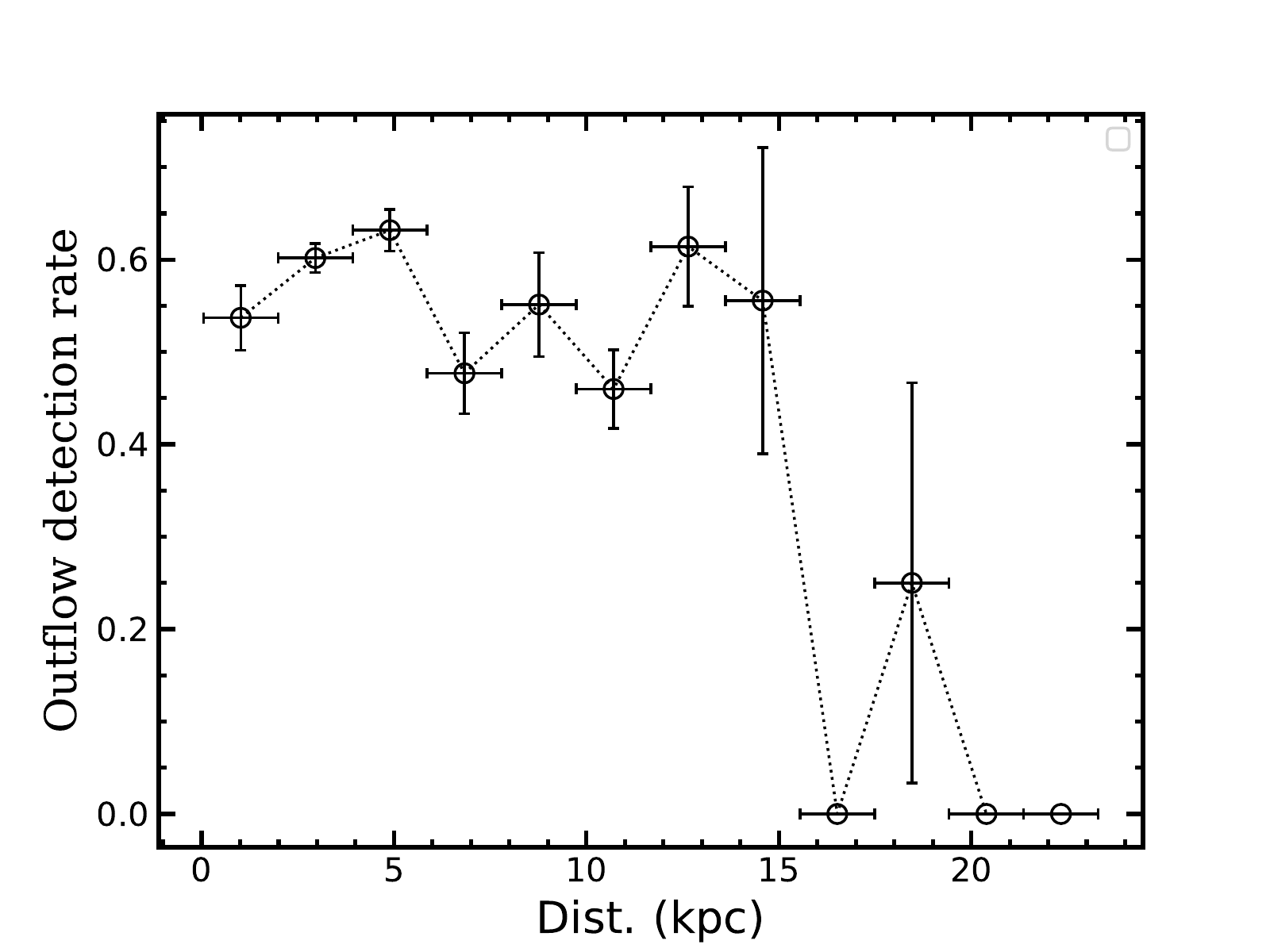} 
 \caption{The Outflow detection rate as a function of clumps' heliocentric distances (Dist.), showing that the detection rates are more or less similar for clumps with Dist. < 14\,kpc, and quickly decrease in clumps with Dist.\,>\,14\,kpc. As <1\% of sources in the sample have Dist.\,>\,14\,kpc, any distance bias would not significantly impact the statistical analysis of the outflow detection in this work. 
 The error bars on the x axis are derived the bin size of distances and y axis are calculated from the standard error of binomial distribution of the detection rate, as shown in Sect.\,\ref{sect:stat_outflow_rates}.}
\label{fig:detection_rate_vs_distance}
\end{figure}

We list the red and blue wing velocities ($\rm V_{min_{b/r}}$,$\rm V_{max_{b/r}}$) and the maximum wing velocity ($\rm \Delta V_{max_{b/r}}$) in Table\,\ref{tab:outflow_wings}, 
where $\rm V_{min_{b/r}}$ and $\rm V_{max_{b/r}}$ refers to the minimum and maximum velocities along the blue or red wings, which indicates the velocity ranges of outflows $\rm \Delta V_{b/r}=|V_{max_{b/r}}-V_{min_{b/r}}|$,
and $\rm \Delta V_{max_{b/r}}$ is the maximum velocity of the blue and red wings relative to the clump velocity (as defined by \cobb), that is, $\rm \Delta V_{max_{b}}=V_{c^{18}o}-V_{min_{b}}$ and $\rm \Delta V_{max_{r}}=V_{max_{r}}-V_{c^{18}o}$, 
 implying the maximum projected velocity of the outflows.
 The wing velocity distributions are presented in Fig.\,\ref{fig:wings_velocity}, which give averages of $\rm \Delta V_{b/r}\sim2.3\,km\,s^{-1}$ and $\rm \Delta V_{max_b/r}\sim4.6\,km\,s^{-1}$.
 A K-S test suggests that there are no significant differences between the velocity ranges of bipolar and unipolar wings, as well as between the red and blue wings. 
 Due to the limitation of sensitivity of this outflow survey, we should bear in mind that the outflow wing velocities are likely to be lower limits. For instance, there are clumps in our outflow sample associated with extremely high-velocity outflows, that is, $\rm \Delta V_{max_{b}}>20\,km\,s^{-1}$ \citep{Choi1993ApJ417624C}, detected in previous work \citep[e.g.,][]{Hervias-Caimapo2019ApJ872200H,Zapata2020ApJ902L47Z,Liu2017ApJ84925L}, and many clumps with very broad SiO\,(2-1) line wings observed by \citet{Csengeri2016AA586A149C}. 
\setlength{\tabcolsep}{2pt}
\def\arraystretch{1.2}
\begin{table}
\centering
\caption {Results of outflow detection rate for clumps in different evolutionary stages.}
\begin{tabular}{p{2.6cm}|C{1.8cm}|C{2.2cm} }
\hline
\hline
 Clumps & Number  & With outflow   \\
 \hline
  Total  & 2052  &   1192\,(58\%)   \\
 \hline
 Quiescent  & 126  &   65\,(51\%)   \\
\hline
 Protostellar  & 322  &   153\,(47\% )  \\
 \hline
 YSO  & 1152  &   656\,(57\% )  \\
 \hline
 MSF  & 428  &   298\,(70\%)   \\
 \hline
  SiO  & 95  &  73\,( 77\%)   \\
\hline
 $\rm CH_{3}OH$ Masers  & 256  &  183\,( 71\%)   \\
 \hline
  $\rm H_{2}O$ Masers  & 180  &  133  \,( 74\%)   \\
  \hline
 $\rm CH_{3}OH$ + $\rm H_{2}O$  & 103  &  76  \,( 74\%)   \\
    \hline
 \uchii\ regions  & 161  &   118\,(73\%)   \\
  \hline
 \uchii\ +$\rm H_{2}O$  & 50  &   39\,(78\%)   \\
 \hline
 \uchii\ +$\rm CH_{3}OH$  & 69  &   57\,(86\%)   \\
 \hline
 \hchii\ regions  & 5  &   5\,(100\%)   \\
 \hline
 \uchii\ $-$Masers  & 74  &   48\,(65\%)   \\
 \hline
\end{tabular}
\label{tab:detection_rate_evotype}
\end{table} 

\begin{figure}
 \centering
   \includegraphics[width = 0.49\textwidth]{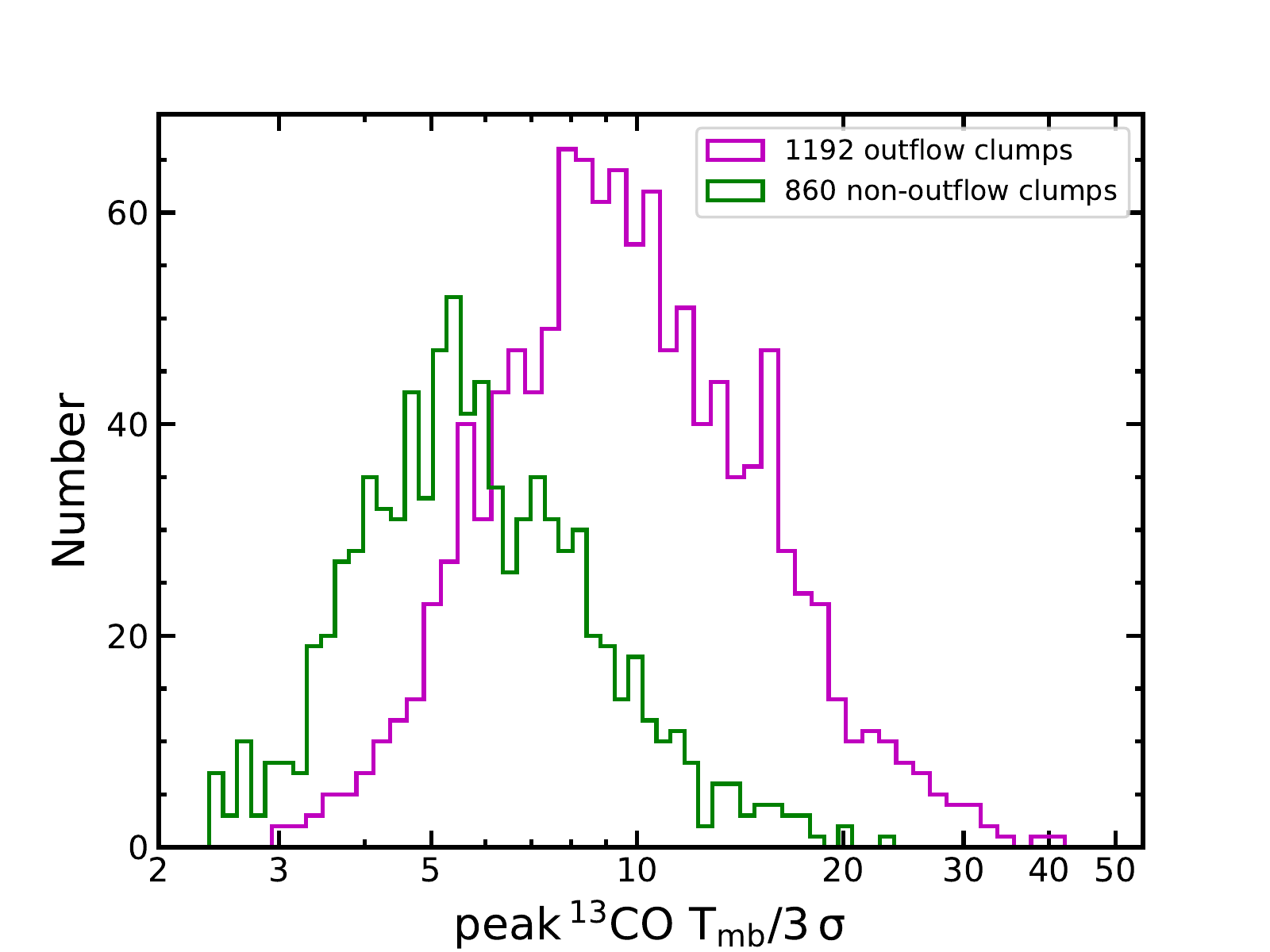} 
 \caption{{The Distribution of the ratio between the peak intensity of the \coaa\ line and the 3$\sigma$ detection level of the corresponding spectrum for the outflow sample (magenta histogram) and the non-outflow sample (green histogram).} 
 }
\label{fig:dis_tpeak_3sigma_ratio}
\end{figure}

\begin{figure}
 \centering
   \includegraphics[width = 0.49\textwidth]{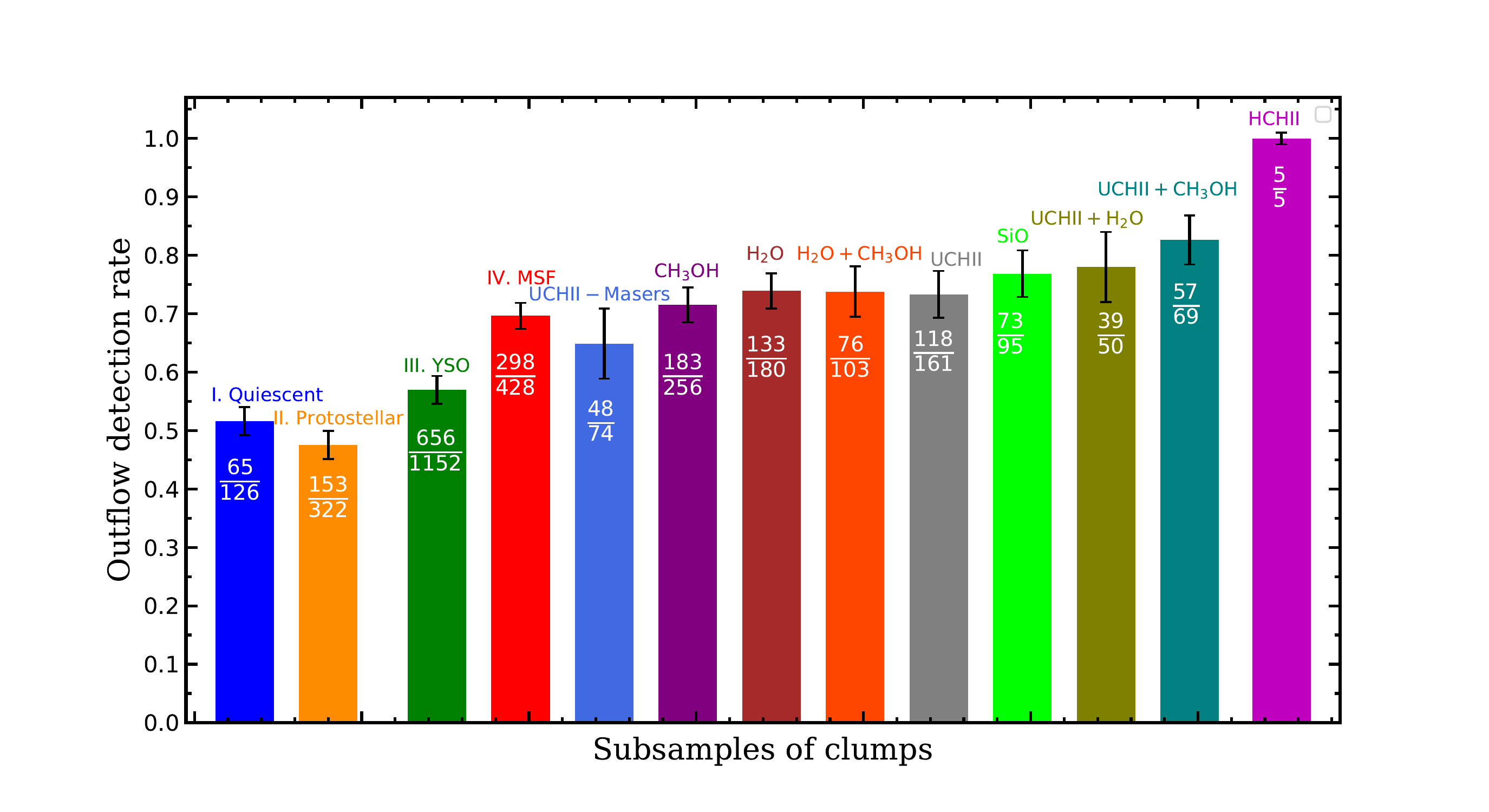} \\ \includegraphics[width = 0.49\textwidth]{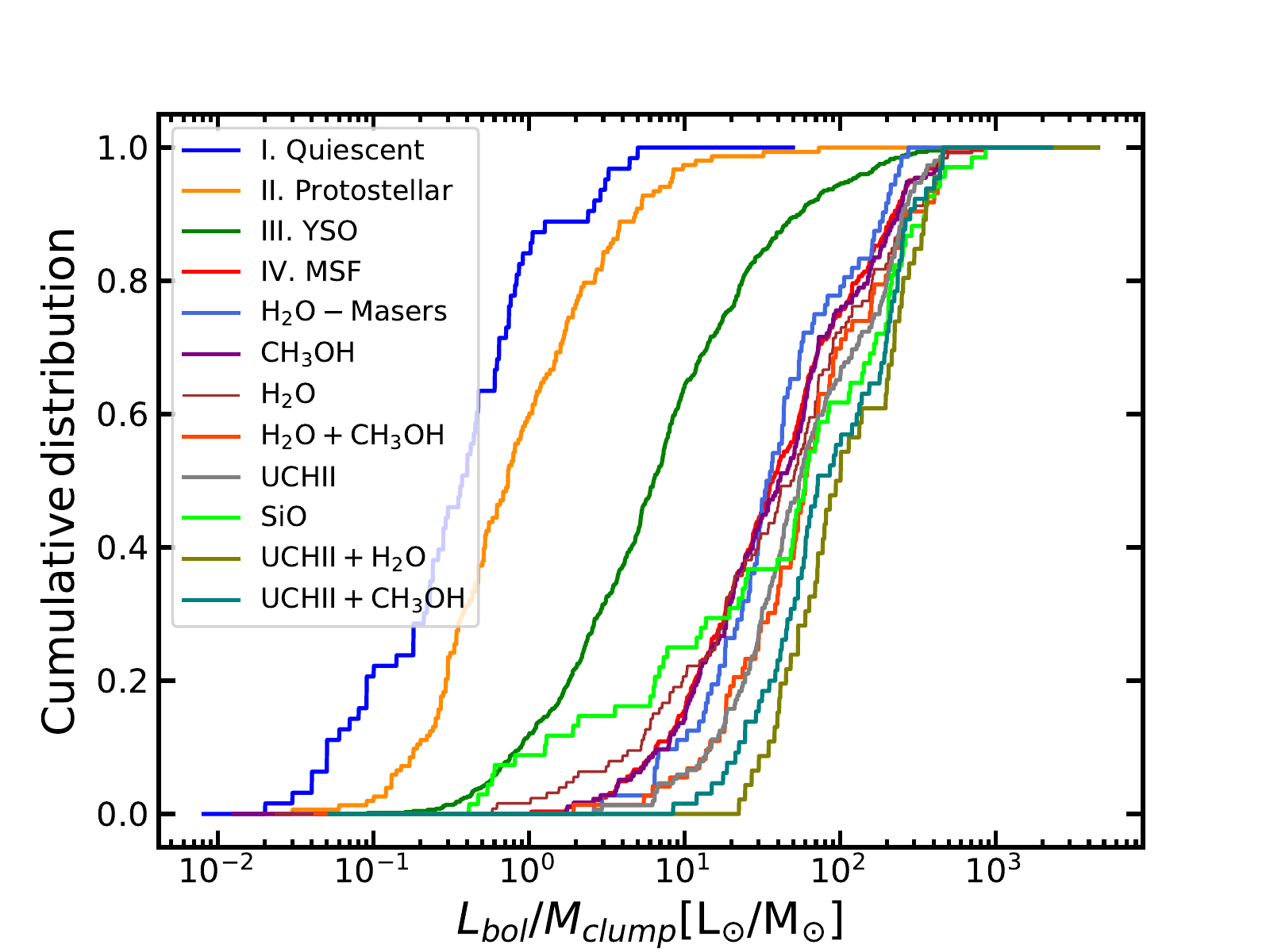} 
 \caption{Distributions of outflow detection rates and luminosity-to-mass ratios for the different subsamples of clumps. Top panel: Outflow detection rate of clumps in different evolutionary stages and associated with different star-forming activities, as presented in Table \ref{tab:detection_rate_evotype}. The error bars of each group are calculated from the binomial errors and the error of the detection rate for clumps with \hchii\ regions is adopted from the binomial errors of the total clump sample. 
 Bottom panel: Cumulative distributions of luminosity-to-mass ratios ($L_{\rm bol}/M_{\rm clump}$) for the subsamples of outflow clumps in the top panel, which indicate the evolutionary stages of these subsamples.}
\label{fig:detection_rate_evotype}
\end{figure}

\subsection{Statistics of outflow detection rates}
\label{sect:stat_outflow_rates}
As outlined in Sect.\,\ref{sect:wings_identification}, if a broader
wing emission in the observed $\rm ^{13}CO$ profile was found, outflows are thought to be detected in the clump. 
We detected high-velocity wings, which we take as being indicative of the presence of outflows toward 1192 of 2052 clumps examined, corresponding to a detection rate of 58$\pm$1\%. Among these, 706 are associated with both red and blue wings ($\sim34\pm1$\%), 230 clumps showing unipolar red wings and 256 clumps having unipolar blue wings (corresponding to 11\% and 13\%, respectively).
The uncertainties of the outflow detection rates in this work and those from the literature are calculated from the standard error of binomial distribution of detection rate, that is, $\rm \sqrt{detection\,rate * (1 - detection\,rate)/(the \,sample\,size)}$. 

To investigate the distance bias for the outflow detection rate, as discussed in Sect.\,\ref{sect:sample}, we present the detection rates as a function of heliocentric distances of clumps in Fig.\,\ref{fig:detection_rate_vs_distance}, showing that the detection rates are more or less similar for clumps with distances < 14\,kpc and sharply decrease in clumps with distances > 14\,kpc. 
This suggests that the detection rate suffers from a distance bias, but only for clumps located further than 14\,kpc; however, <1\% of the sources have distances > 14\,kpc in the 2052 total clump sample, and the 1192 outflow sample. Therefore, the distances bias for the outflow identification would not be significant and the systematic analysis of the detection rate is valid. 

Our overall detection rate of $58\pm1\%$ (1192/2052) is derived from the largest sample of clumps (2052), using \coaa\ lines. 
It is comparable to the outflow frequency in many previous studies using CO lines,  such as \citealt{Zhang2001ApJ552L,Zhang2005ApJ625} (57$\pm$6\%) for 69 $IRAS$ point sources (rms$\sim$0.05\,K), and \citealt{wu2010ApJ720W} (59$\pm$12\%) for 17 maser sources (rms$\sim$0.1\,K). 
It is larger than the detection rate in  \citealt{Codella2004AA615C} (39\%$-$50\%) for 136 \uchii\ regions using \coa\ (rms$\sim$0.5\,K), and \citealt{Li2018ApJ867167L} (20\%) toward a sample of 770 clumps using \coac\ data (rms$\sim$2\,K) from the COHRS survey \citep{Dempsey2013ApJS2098D}. 
 It is, however,  slightly smaller than Paper\,I (69$\pm$3\%) for 325 massive clumps using \coab\   data from the CHIMPS survey ($\rm rms\sim 0.6\,K$), and \citealt{Maud2015MNRAS645M} (66$\pm$5\%) for 89 massive young stellar objects and compact \hii\ regions (rms$\sim$0.6\,K). 
The detection rates for  clumps associated with  massive star-forming (MSF) activities are $\sim 70$-$80\%$ (see Table\,\ref{tab:detection_rate_evotype}), which are also slightly smaller than the work of targeted observations toward tracers of MSF, such as \citealt[]{Xu2006AJ13220X} (88$\pm$11\%) for eight methanol masers using \coae\ (rms$\sim$0.2\,K), \citealt{deVilliers2014MNRAS444} (100\%) for 54 methanol masers using \coab\ (rms$\sim$0.36\,K), \citealt{Lopez_Sepulcre2009AA499L} (100\%) for 11 very luminous massive YSOs using \coaf\ (rms$\sim$0.3\,K), \citealt{Shepherd1996ApJ457S} (90$\pm$3\%) for 94 high-mass star-forming regions  using CO\,(1-0) (rms$\sim$0.02\,K), \citealt{Beuther2002AA383B} (81$\pm$8\%) for 26 high-mass star-forming regions using CO\,(2-1), and \citealt{Lopez_Sepulcre2011AA526L2L} (88$\pm$4\%) for 57 high-mass molecular clumps  using SiO (rms$\sim$0.01\,K).
We note that the detection rates from these are obtained from limited samples, with small numbers of targets in a particular stage, class, or category, compared to this study.
Given the variations in the detection  sensitivities and lines used, the differences in the detection rate between the studies are not considered to be significant.
Thus an overall detection rate of 58\% can be considered in reasonable agreement with the previous surveys.

The outflow detection rate is obviously a lower limit and one may wonder what fraction of the non-outflow sources might be associated with outflow wings if observed with better sensitivity.
In this respect, it is interesting to consider the distribution of the ratio between the line peak and the 3-$\sigma$ detection limit of outflow wings for the two samples, that is, the outflow sample and the non-outflow sample, as illustrated in Fig.\,\ref{fig:dis_tpeak_3sigma_ratio} (a similar approach was used by \citet{Palagi1993AAS101153P}; see their fig.14). 
The K-S test of the ratio between the two samples suggests that they are significantly different, with $p$-value\,$\ll$\,0.0001. 
The fact that the two histograms peak at different values of that ratio proves that the intensity of the line wings, $\rm T_{wing}$, must depend somewhat on the line peak, $\rm T_{peak}$. In fact, if the two
were independent of each other, apart from the obvious condition $\rm T_{wing}<T_{peak}$, the two distributions should largely overlap, because the criterion used to discriminate between the two samples (the detection of line wings) would be almost unrelated to the parameter used to make the histograms (the line peak intensity).

We conclude that the ratio between the line peak and wing intensities, $\rm T_{peak}/T_{wing}$, is not a random number >1  for outflows to become detectable 
but must span a limited range of values. In order to clarify this concept, one may consider the extreme (unrealistic) example
where such a ratio is exactly equal to N. In this case the distribution of non-outflow sources would be strictly
confined to $\rm T_{peak}/3\sigma<N$, whereas that of outflow sources would have only $\rm T_{peak}/3\sigma>N$.
The real world is clearly more complicated, and one expects this ideal scenario to be "spoiled" by various effects.
One is the presence of truly non-outflow sources (i.e., lines intrinsically devoid of wings), which broaden toward $\rm T_{peak}/3\sigma>N$
the distribution of the so-called non-outflow sources; this can explain the long tail of the non-outflow histogram
in Fig.\,\ref{fig:dis_tpeak_3sigma_ratio}, extending up to $\rm T_{peak}/3\sigma=20$. Furthermore, $\rm T_{peak}/T_{wing}$ depends upon
many source-dependent physical factors (including the inclination of the outflow with respect to the line of sight) that
are bound to make the separation between the two distributions much less sharp than in our illustrative example.

Despite all these caveats, the basic idea remains valid and
the shift between the two histograms in Fig. 7 must convey the information on the typical ratio
between line peak and wing intensities. Based on the previous discussion, we conclude that $\rm T_{peak}/T_{wing}$ should lie roughly
between the peaks of the two histograms and thus range approximately between 4 and 10. As a consequence, only the
sources in the non-outflow sample with $\rm T_{peak}/3\sigma>10$ are in all likelihood not associated with outflows if outflows are not aligned with the plane of the sky. 
The rest of the objects classified as
non-outflow should contain a considerable fraction of outflow sources in disguise.

\begin{figure*}
\centering
\begin{tabular}{ccc}
 \includegraphics[width = 0.32\textwidth] {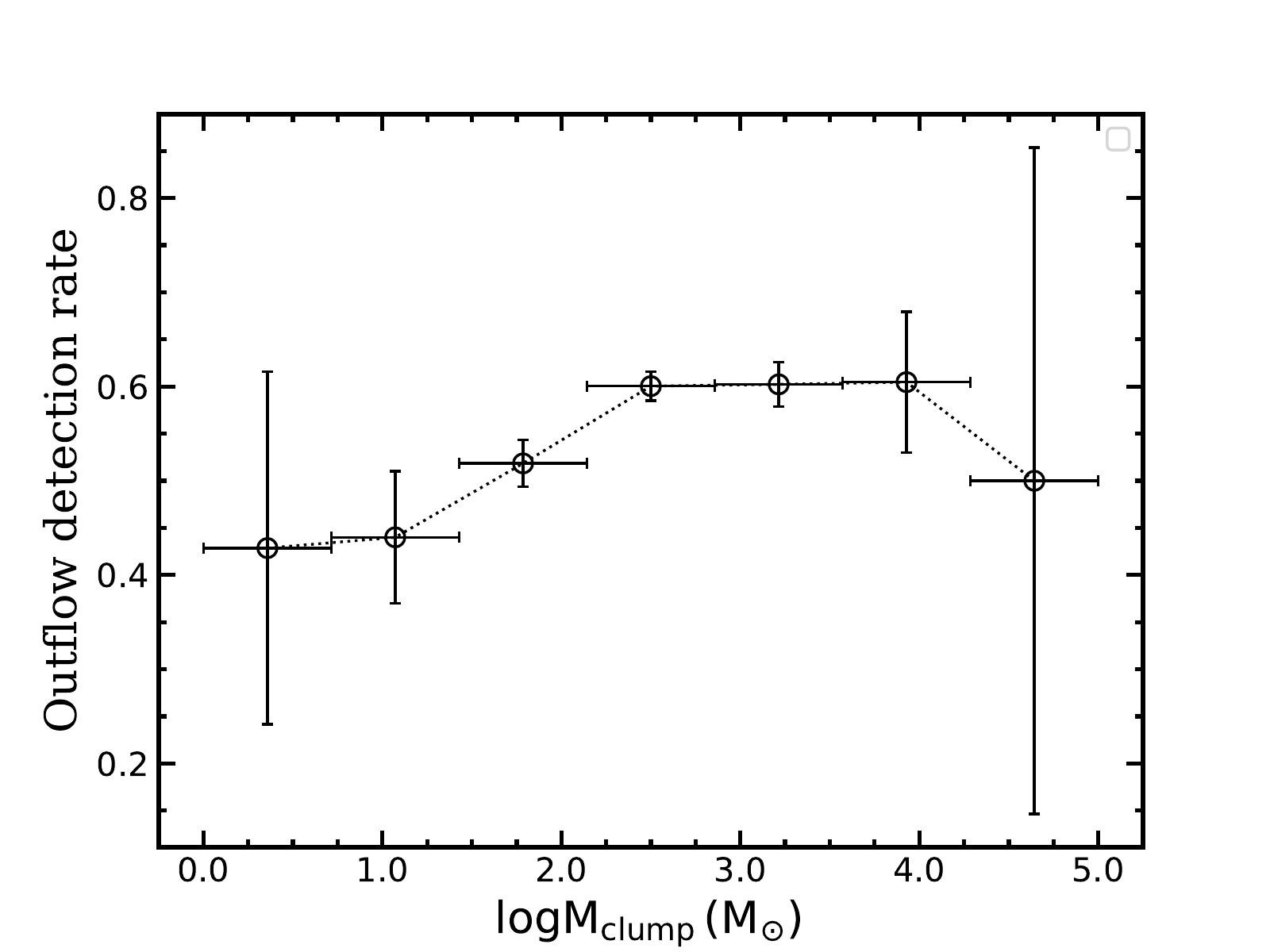} & \includegraphics[width = 0.32\textwidth] {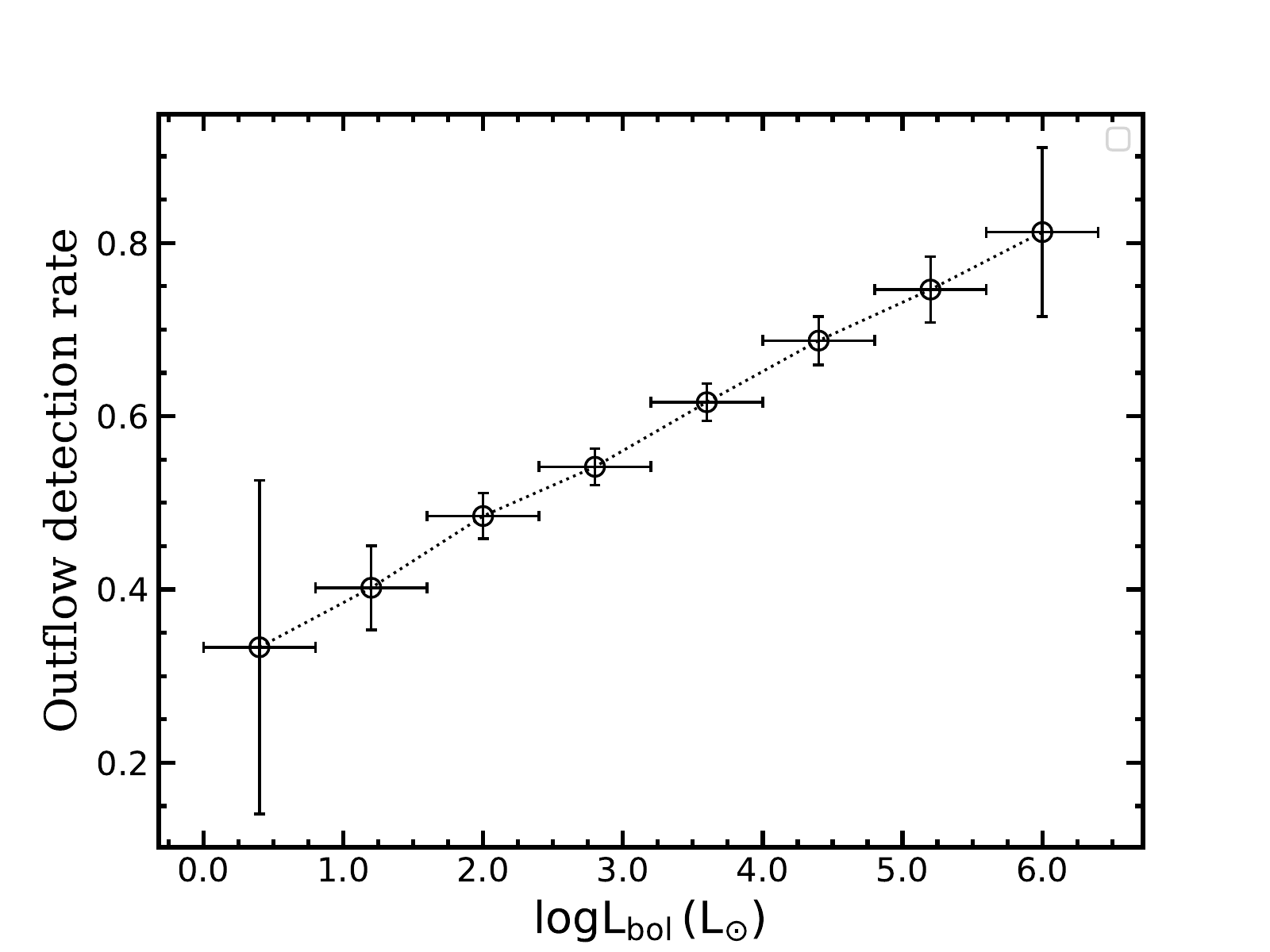} & 
 \includegraphics[width = 0.32\textwidth] {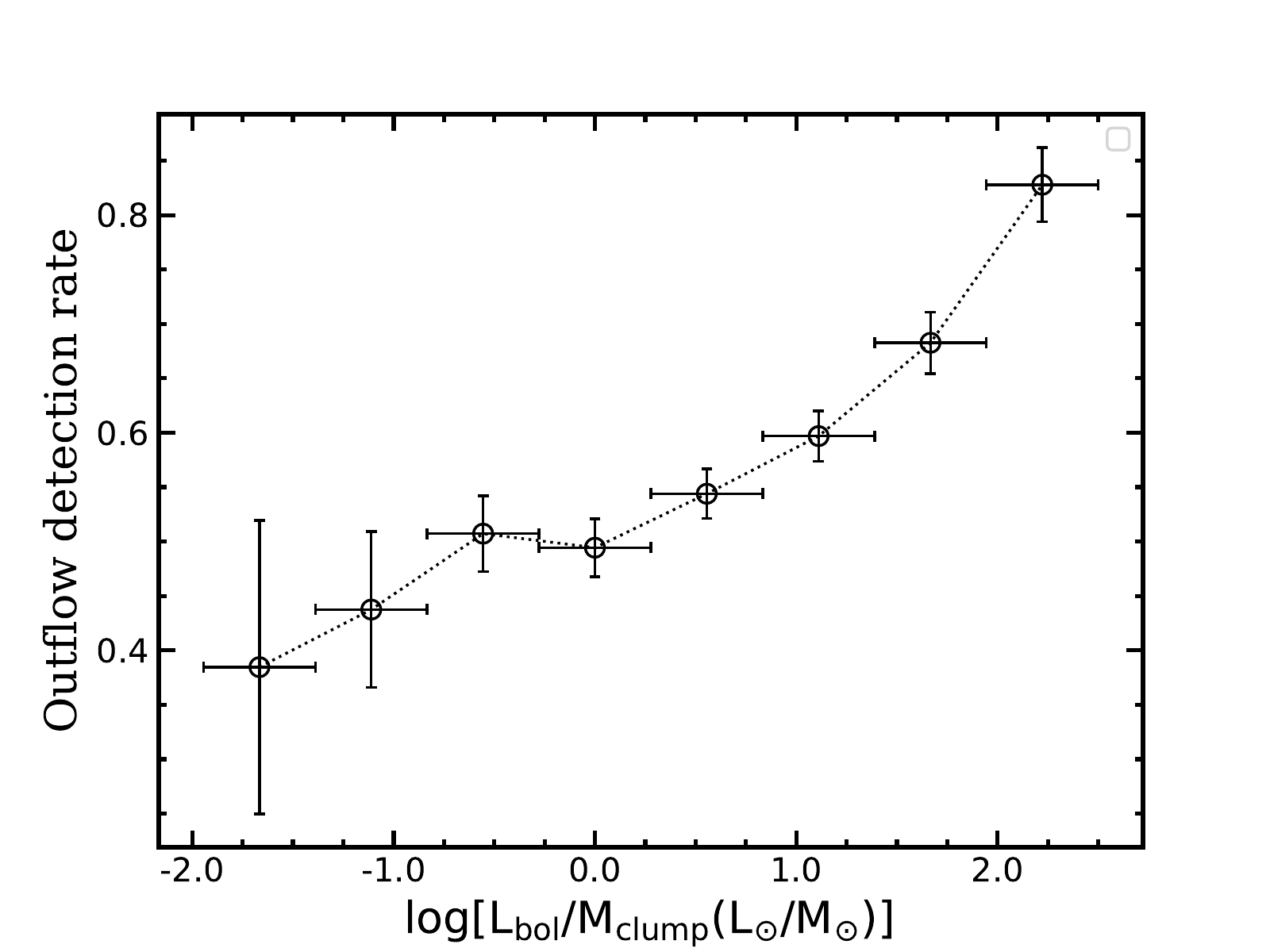} \\
 (a)& (b) & (c) \\
 \includegraphics[width = 0.32\textwidth] {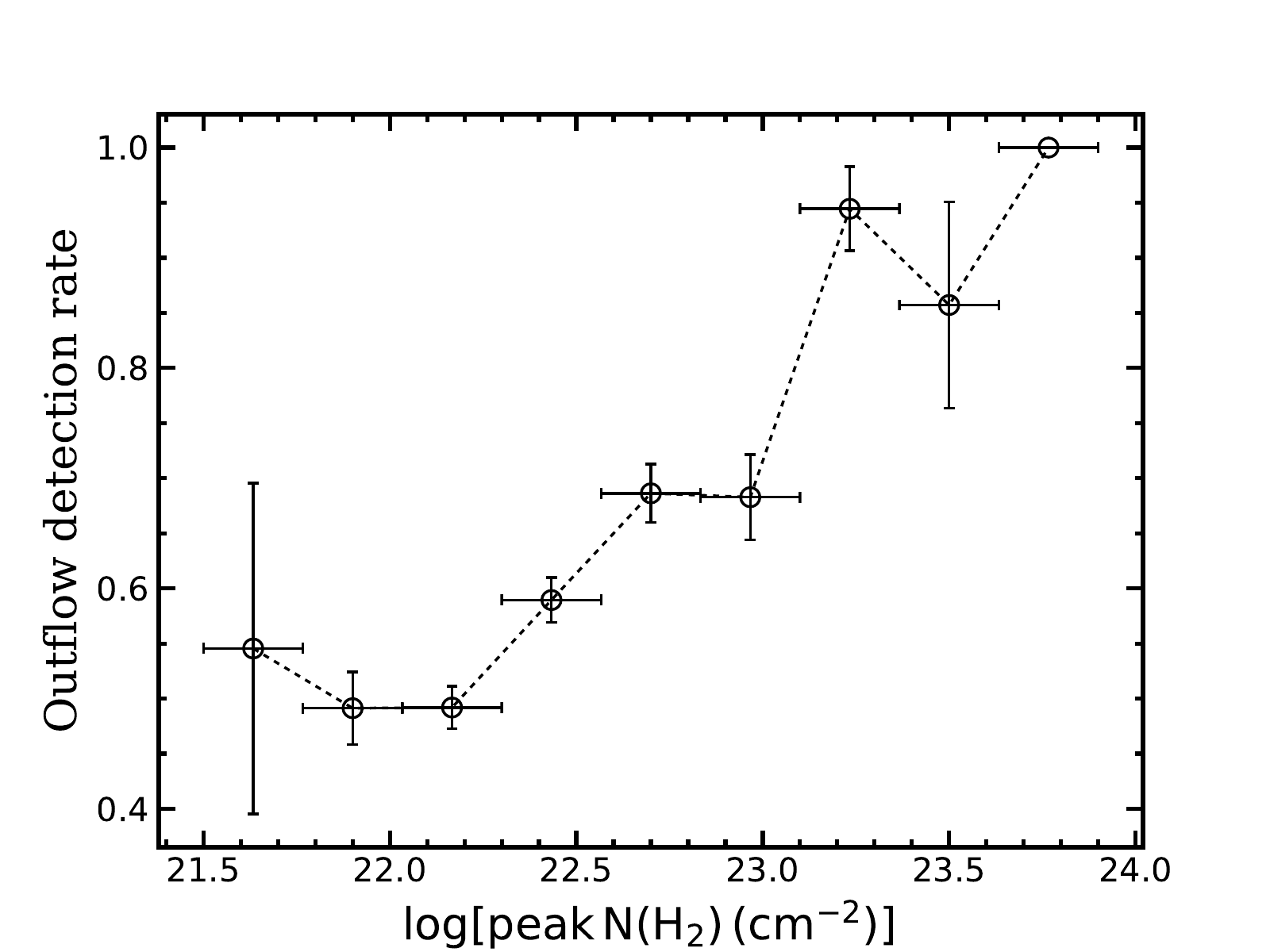} & \includegraphics[width = 0.32\textwidth] {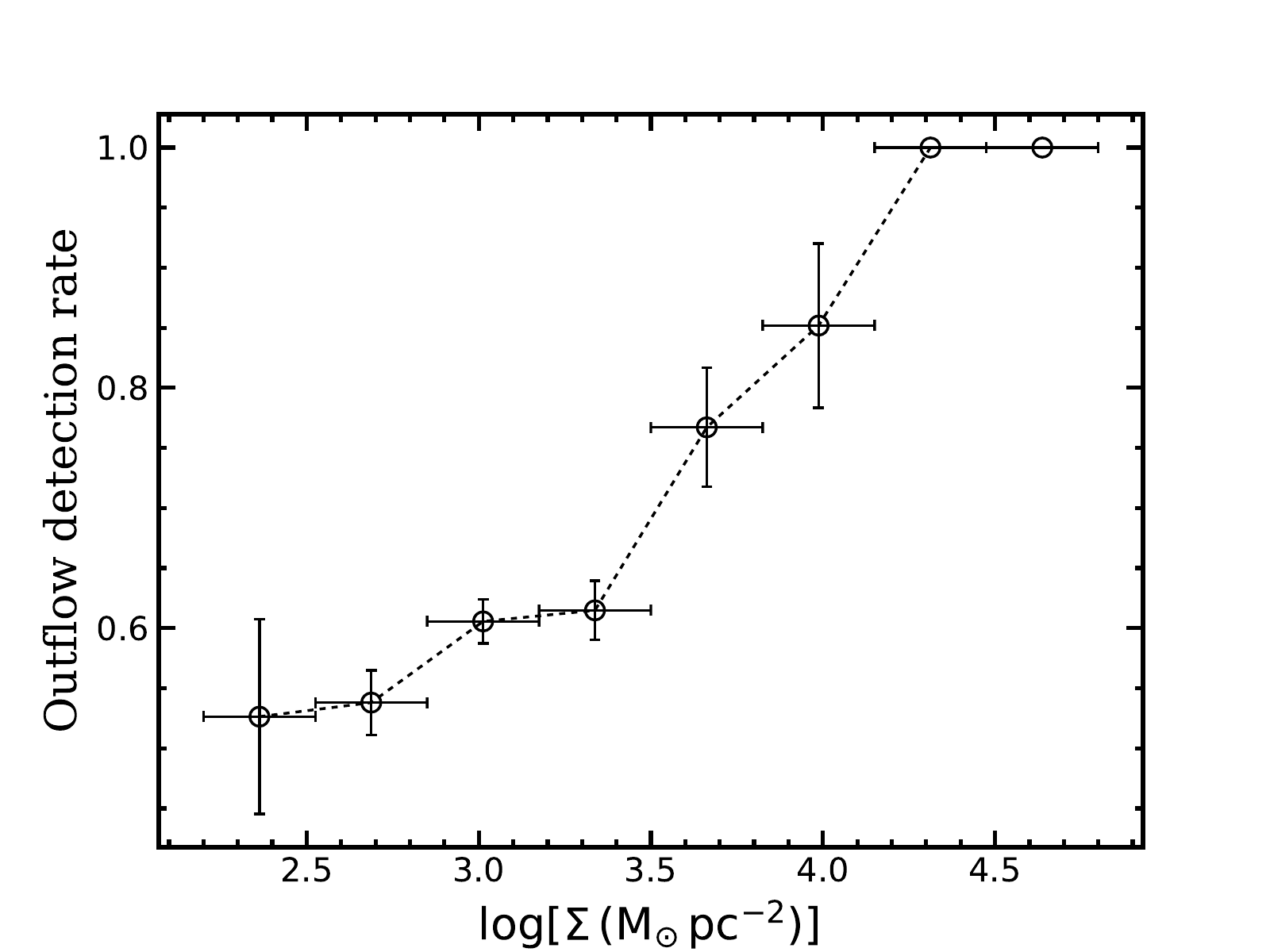} & 
 \includegraphics[width = 0.32\textwidth] {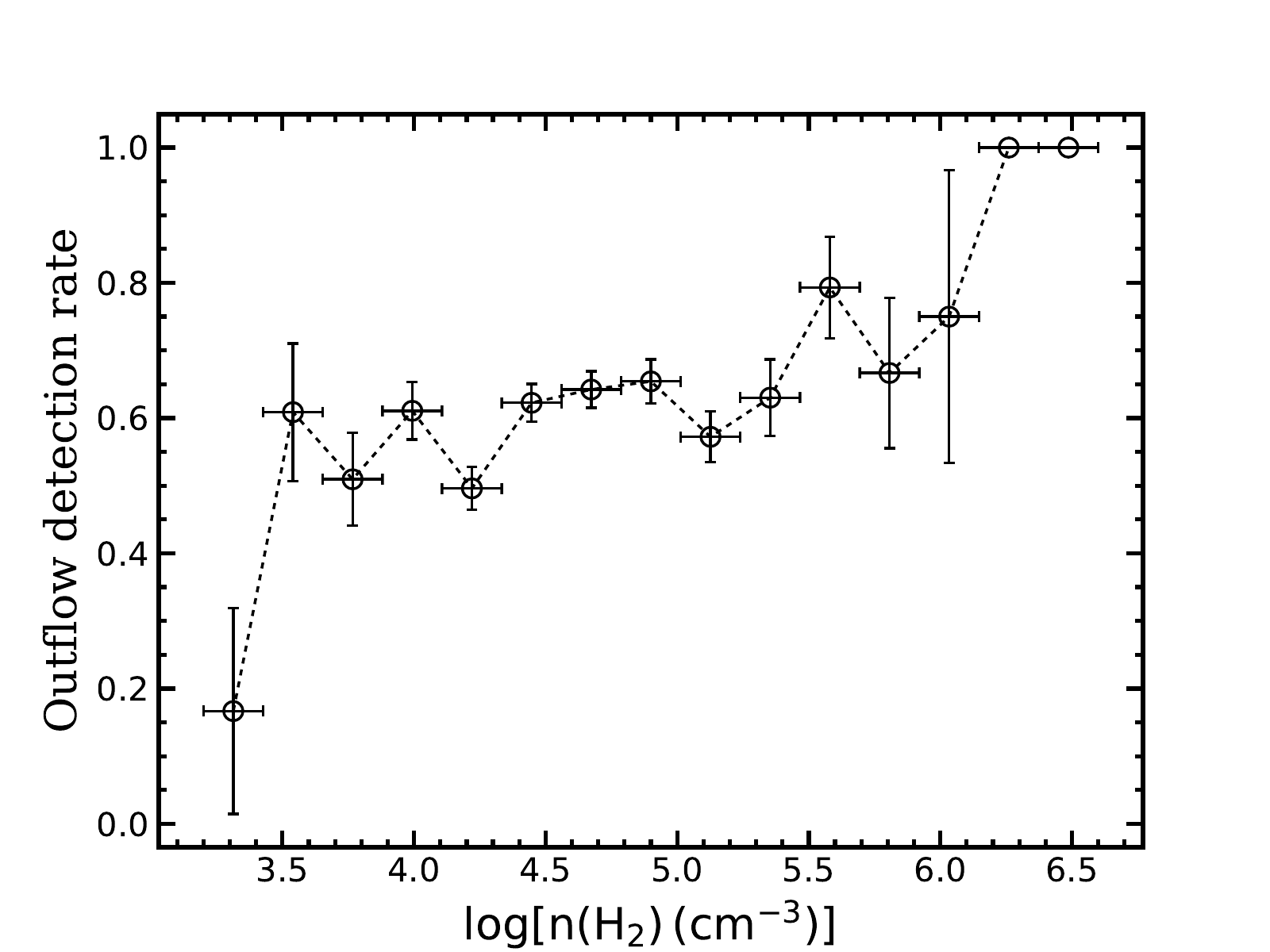} \\
 (d) & (e) & (f) \\
\end{tabular}
 \caption{Outflow detection rate against the physical properties of clumps. Top left (a,b,c) to bottom right (d,e,f): Detection rate as a function of clump mass $\rm M_{clump}\,(M_{\odot})$, 
 bolometric luminosity of the central object $\rm L_{bol}/L_{\odot}$, 
 luminosity-to-mass ratio $\rm L_{bol}/M_{clump}\,(L_{\odot}/M_{\odot})$, the peak $\rm H_2$ column density [$\rm peak\,N({H_2})/cm^{2}$], the mean mass surface density $\rm \Sigma\,(M_{\odot}/pc^{2})$, and the mean volume $\rm H_2$ density $\rm n({H_2})/(cm^{3})$ of clumps in logarithmic scales. The error bars on the x axis and y axis are determined following the same method in Fig\,\ref{fig:detection_rate_vs_distance}.
 }
\label{fig:detection_rate_vs_param}
\end{figure*}

\subsection{The evolutionary trends of outflow detection rates}
\label{sect:evolution_detection_rate}
The evolutionary trend of outflow detection rates was discussed in Paper\,I, on the basis of the four evolutionary stages of ATLASGAL clumps as classified by \citet{koenig2017AA599A139K} and \citet{Urquhart2018MNRAS4731059U}, the earliest quiescent clumps (i.e., a starless or pre-stellar phase that is $\rm 70\mu m$ weak), protostellar (i.e., clumps that are $\rm 24\mu m$ weak but far-infrared bright), YSO-forming clumps (YSO clumps; i.e., $\rm 24\mu m$-bright clumps), and massive star-forming clumps (MSF clumps; i.e., $\rm 24\mu m$-bright clumps with a tracer of massive star formation). 
\citet{Yang2018ApJS235Y} found that outflow activity becomes much more common as clumps evolve from the earliest quiescent clumps (2/4; $50\pm25\%$), to protostellar clumps (10/19; $53\pm11\%$), to  YSO-associated clumps (105/171; $61\pm4\%$) to MSF-associated clumps (102/125; $82\pm3\%$). 
The trend of increasing detection rate with the evolutionary stage of clumps is also found for  our significantly larger sample here, presented with very comparable detection statistics: 
from quiescent clumps (65/126; 51$\pm$1\%), to protostellar clumps (153/322; 47$\pm$1\%), to YSO clumps (656/1152; 57$\pm$1\%), and to MSF clumps (298/428; 70$\pm$1\%) \footnote{24 clumps that have not yet been classified into the four evolutionary stages.}. 
Among the four evolutionary stages, the outflow detection rates remain almost unchanged at the first two stages, show an obvious increase in the third YSO phase, 
and reach the peak of 70$\pm$1\% for clumps in the MSF stage. 
 This increasing trend of detection rate among the four evolutionary stages of clumps is also indicated in Fig.\,\ref{fig:detection_rate_evotype}.

For clumps in the MSF stage, that is, the mid-infrared $\rm 24\,\mu m$-bright clumps associated with \hii\ regions and MYSOs by the RMS survey \citep{lumsden2013}, water masers by HOPS survey \citep{Walsh2011MNRAS1764W}, and methanol masers identified by the MMB survey (see \citealt{urquhart2013a,urquhart2013b,urquhart2014c,urquhart2015_methanol} for details), the probability to have an outflow is expected to rise, as discussed in Paper\,I. 
In this work, the outflow detection frequency of outflows increases  to 72\% (240/333) in clumps  with associated maser emission (6.7\,GHz $\rm CH_{3}OH$ and/or 22.2\,GHz  $\rm H_{2}O$  \citep{Billington2020MNRAS4992744B}), compared to $\sim$57\% in those without maser associations. 
The outflow detection rate is quite high for clumps associated with both $\rm CH_{3}OH$ and $\rm H_{2}O$ masers (74$\pm$4\%, 76/103). 
There is also a high detection rate toward clumps associated with SiO (5-4,3-2,2-1) emission, which is a shocked gas tracer (76.8\%; 73/95), based on the SiO data collected from the SEDIGISM survey and the literature \citep{Harju1998AAS132211H,Csengeri2016AA586A149C,Stroh2019ApJS24425S}, as marked in Table\,\ref{tab:clump_properties}. The SiO emission is expected to be enhanced at the edges of the outflow cavity where material in the outflow is colliding with the ambient medium and so the high correlation provides strong support that a molecular outflow is present.

The detection rate toward the 161 clumps associated with \uchii\ regions is 73\% (118/161) \citep{urquhart2013b,Kalcheva2018AA615A103K}. 
The outflow wing detection rate in \uchii\ regions can be as high as 80\% (70/87) for clumps associated with \uchii\ regions and masers ($\rm H_{2}O$: 39/50; $\rm CH_{3}OH$: 57/69), and it drops to 65\%\,(48/74) for clumps with \uchii\ regions without any maser emission.
The detection fraction can even rise to 100\% (5/5) for clumps associated with hyper-compact (HC) \hii\ regions (\citealt{Yang2019MNRAS4822681Y,Yang2021AA645A110Y}). However, we note that the sample size is small. 
This suggests that the outflow detection rate appears to peak in the pre-\uchii\ stage (i.e., \hchii\ regions, maser-associated \uchii\ regions), as stated in Paper\,I, and then it starts to decrease in the late \uchii\ regions phase (no-maser-associated \uchii\ regions). 
 Alternatively, the increase in detection rate might also be the result of the increasing probability of detection due to the increase in energy and poorer collimation of the outflow with age. This can be supported by the bottom panel of Fig.\,\ref{fig:detection_rate_evotype}, which shows that the higher detection rate is associated with clumps with a larger luminosity-to-mass ratio ($L/M$). As the outflow energy is positively related to the $L/M$ \citep{Wu2004AA426W,deVilliers2014MNRAS444,Maud2015MNRAS645M,Yang2018ApJS235Y}, the clumps with higher $L/M$ produce stronger outflows that are easier to detect.

The growth trend of the outflow detection rate is also seen in the increasing physical properties of clumps. 
These physical parameters are the clump mass ($\rm M_{clump}/M_{\odot}$), bolometric luminosity of central objects ($\rm L_{bol}/L_{\odot}$), luminosity-to-mass ratio $\rm (L_{bol}/M_{clump}\,[L_{\odot}/M_{\odot}]$), the mean mass surface density ($\rm \Sigma\,[M_{\odot}\,pc^{-2}]$) and mean $\rm H_2$ volume density ($\rm n(H_{2})/cm^{3}$) within the 50\% contour, as well as the peak $\rm H_2$ column density of the clumps [peak\,$\rm N(H_{2})/cm^{2}$]. 
For each parameter, we divide the total clump sample into several bins covering the minimum to maximum values given in Table\,\ref{tab:summary_param}, with equal bin widths. We can then compare the detection rates as a function of the physical properties of the clumps. 
The detection rate as a function of these parameters are presented in Fig.\,\ref{fig:detection_rate_vs_param}.  
The detection rates rise with an increase in $\rm L_{bol}$, $\rm L_{bol}/M_{clump}$, and densities, while approaching $\sim 80\%$ for clumps with high values of these parameters. 
 This can be seen from the cumulative distributions of these parameters between the outflow sample and the non-outflow sample presented in Fig.\,\ref{fig:cdf_distribution_param_outflow}, showing that the outflow sample has higher values of these properties than the non-outflow sample. 
This indicates that more luminous, dense, and evolved sources show a much higher outflow detection fraction, in agreement with Paper\,I. 
We also note that the outflow detection rate shows a small increase as a function of clump mass, which means that the strong correlation between $\rm L_{bol}/M_{clump}$ and outflow detection rate is almost entirely driven by luminosity. 
This is also consistent with \citet{Urquhart2018MNRAS4731059U} who find clump mass is independent of evolutionary stage. 
Also, from the correlation with these physical parameters in Fig.\,\ref{fig:cdf_distribution_param_outflow}, the outflow detection rate is found to be strongest correlated with the clump luminosity, with a correlation coefficient of $\rho=$0.999 and p$-$value$\ll0.001$. 
Therefore, the luminosity is most closely related to the outflow detection rate, which is consistent with Paper\,I that also found the strongest correlation between outflow properties and clump luminosity. 
As stated in Paper\,I, we find that there are a few clumps at a later stage of evolution,  with large values of $\rm L/M$ and $L$,  which show no evidence for outflow wings; this may be due to outflow geometry, the complexity of the CO emission
\citep[e.g.,][]{Zhang2001ApJ552L}, interactions of the sources within the clumps below our resolution \citep[e.g.,][]{Codella2004AA615C}, or external winds and/or shocks \citep[e.g.,][]{Maud2015MNRAS645M}.

\section{Discussion}
\label{sect:discussion}

\subsection{What could cause the variation in  detection rate?}
\label{sect:the_common_of_outflow}

Based on the large clump sample in this work, the detection frequency of outflows is found to increase as a function of their four evolutionary stages, namely, from the youngest quiescent (51\%), to protostellar (47\%), to YSO (57\%), and to MSF clumps (70\%), as seen in Paper\,I.  
This is supported by the outflow detection rates rising with the increase in the luminosity-to-mass ratio $\rm L_{bol}/M_{clump}$, as shown in Fig.\,\ref{fig:detection_rate_evotype} and panel (c) of  Fig.\,\ref{fig:detection_rate_vs_param}. 
This ratio has been widely studied and used in the literature as an evolutionary tracer \citep[][and reference therein]{Urquhart2018MNRAS4731059U}, but \citet{Pitts2019MNRAS305P} have shown that it is physically the same as clump-averaged $\rm T_{dust}$, and indeed, does not simply trace clump evolution at lower clump densities than accessed by ATLASGAL. Nevertheless, the outflow detection rates clearly vary with $\rm L_{bol}/M_{clump}$ as shown in Fig.\,\ref{fig:detection_rate_vs_param}. 
The variation in outflow detection is also seen in the MSF stages, ranging from 70\% to $\sim$100\% for clumps associated with different tracers of massive star formation, as shown in Table\,\ref{tab:detection_rate_evotype} and Fig.\,\ref{fig:detection_rate_evotype}. 
In total, the outflow detection rates are variable for different stages of clumps.
\begin{figure}
 \centering
   \includegraphics[width = 0.49\textwidth]{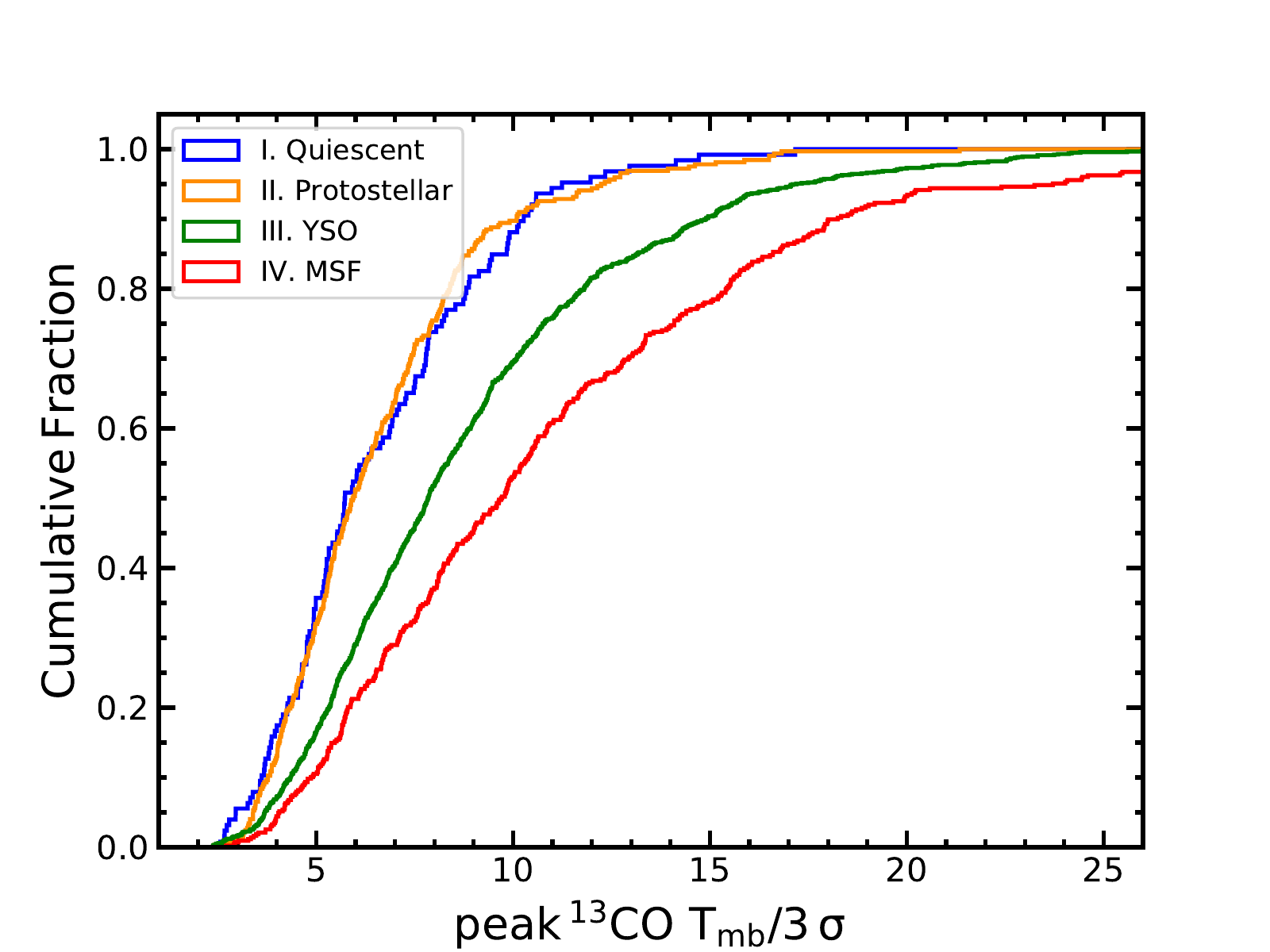} \\
\includegraphics[width = 0.49\textwidth]{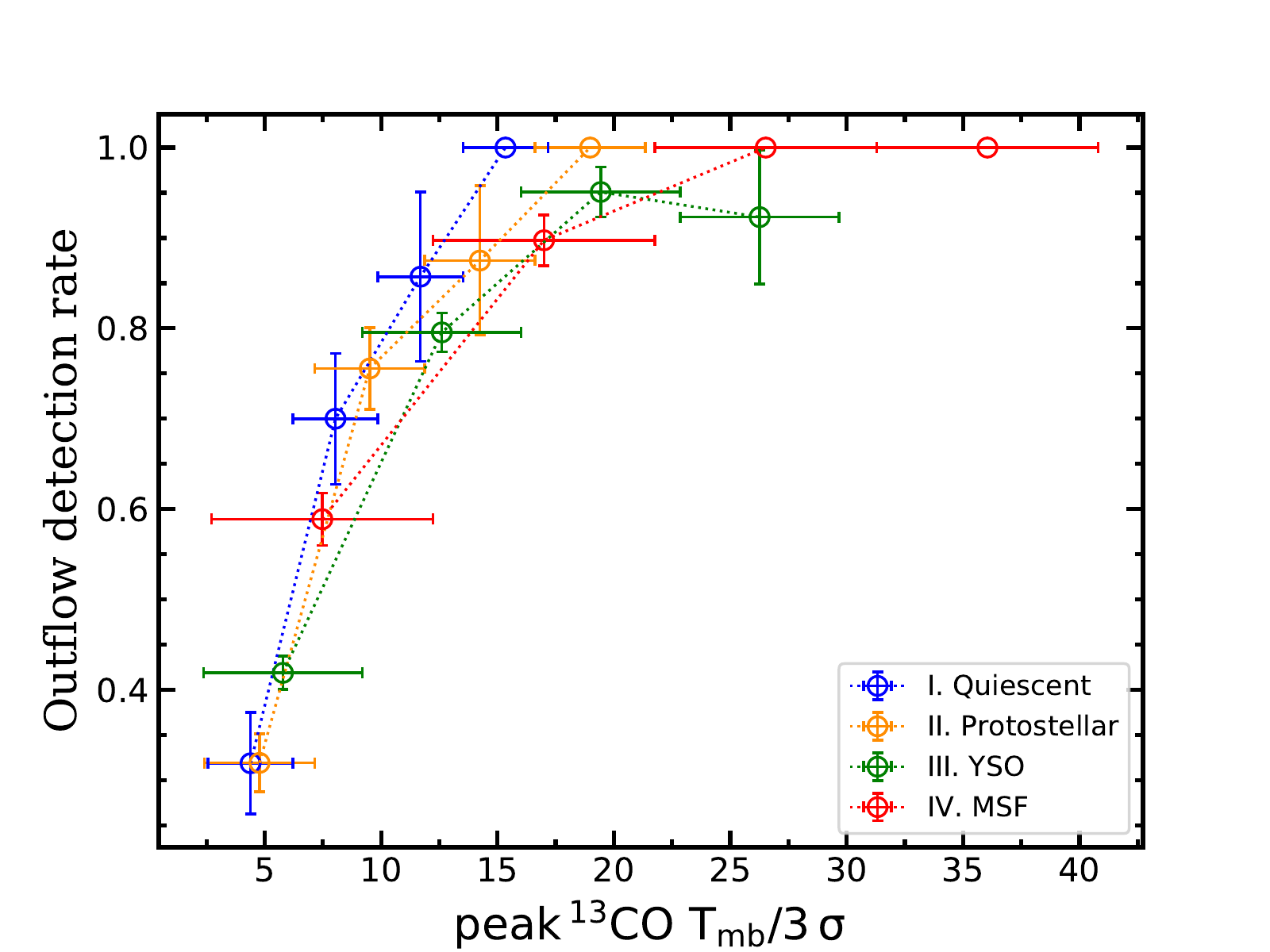} \\
 \caption{ Distributions of the peak to $3\sigma$ ratio and its relation with outflow detection rate for clumps in different stages. Top panel: the Cumulative distributions of the ratio between the peak of \coaa\  and 3$\sigma$ noise level of each corresponding spectrum [peak \coa\ $\rm T_{mb}/3\sigma$] for clumps in the four evolutionary stages in the total 2052 sample. Bottom panel: Outflow detection rate as a function of the [peak \coa\ $\rm T_{mb}/3\sigma$].
 This indicates that outflow wings are stronger for clumps in the late evolutionary stages and thus easier to detect, compared to clumps in the early evolutionary stages. The detection rates of outflows are likely to be similar for clumps in every evolutionary stage. 
 The error bars on the x axis and y axis are determined following the same method in Fig\,\ref{fig:detection_rate_vs_distance}.}
\label{fig:cdf_tpeak_13co}
\end{figure}

What could be the reason responsible for the variations in outflow detection rate for the different stages of clumps?
As discussed in Sect.\,\ref{sect:evolution_detection_rate}, 
the outflow detection rates show the strongest correlation ( $\rho=$ 0.999 and p$-$value\,$\ll\,0.001$) with the clump bolometric luminosity. 
The clump luminosity strongly increases as a function of the four evolutionary stages of clumps and clump mass is found to be independent of evolution \citep{Urquhart2018MNRAS4731059U}.
Therefore, the correlation between outflow detection rate and $\rm L_{bol}/M_{clump}$
may result from the strong increase in $\rm L_{bol}$ as the clump evolves.
Additionally, the variation in outflow detection rates of the different stages of clumps may be because they have different luminosities. 

Furthermore, the clump outflow energy is found to increase as a function of clump luminosity   \citep[e.g.,][]{Wu2004AA426W,deVilliers2014MNRAS444,Yang2018ApJS235Y}. 
Clumps in the late stages with higher luminosities \citep{Urquhart2018MNRAS4731059U} are likely to be associated with stronger outflows (Paper\,I) that are easily detected, resulting in high outflow detection rates. 
The lower detection rates of outflows in the three earlier stages (i.e., quiescent, protostellar, and YSO stages) are, therefore, possibly due to the fact that outflows are weak in these phases and thus difficult to detect. 
It is then possible that the increase in detection rate with the evolutionary stage is due to an increase in detectability rather than a lower number of outflows clumps being present. 

This hypothesis is supported by the top panel of Fig.\,\ref{fig:cdf_tpeak_13co}, showing that the clumps in the later stages have higher values of the peak to noise ratio ([peak \coaa\ $\rm T_{mb}/3\sigma$]), that is, the detectability of outflow wings, compared to the clumps in earlier stages.
In addition, as presented in the bottom panel of Fig.\,\ref{fig:cdf_tpeak_13co}, outflow detection rates increase as the increasing values of [peak \coaa\ $\rm T_{mb}/3\sigma$] for clumps in each of the four evolutionary stages. 
For clumps with high values of [peak \coaa\ $\rm T_{mb}/3\sigma$], the outflow detection rates appear to be similar for the four stages. 

As discussed above, the variation in detection rates of different stages of clumps may be due to the difference in outflow detectability, and the existence of outflows are possibly similar for each of the four evolutionary stages of clumps. 
Therefore, the outflow activities are suggested to be switched on in the earliest quiescent stage (i.e., 70\,$\mu$m weak or dark clumps) rather than the late stages, and present through all four stages of evolution. 
Furthermore, the detection of outflows in these 70\,$\mu$m dark ATLASGAL clumps suggest that they are not quiescent, and the absence of 70\,$\mu$m emission is not a robust indicator of starless and/or pre-stellar cores.  
This is also supported by the detection of outflows at a high resolution ($\sim 3\arcsec$) by \citet{Li2019ApJ886130L}  from a sample of seven 70\,$\mu$m dark ATLASGAL clumps.



\subsection{Implication of the high detection rates of outflows}
\label{sect:implication_of_high_detections}

Based on the results outlined above, we can find that the probability of detecting outflow wings is very high for massive and luminous clumps at later evolutionary stages, approaching $\sim$80\% when $\rm \log[L_{bol}/M_{clump}\,(L_{\odot}/M_{\odot})]>1.7$\,(94/117), $\rm \log[L_{bol}/L_{\odot}]>5.34$\,(41/50), as shown in Fig.\,\ref{fig:detection_rate_evotype} and Fig.\,\ref{fig:detection_rate_vs_param}.
We may miss some outflows due to disadvantageous inclination angles, but the high detection rate indicates that outflows are common toward these luminous clumps. 

The high detection rate of outflows would be expected if high-mass stars are formed via an accretion disk, namely, a scaled-up version of low-mass star formation \citep[e.g.,][]{Zhang2001ApJ552L,Kim2006ApJ643}. 
Recently, a growing number of disk candidates have been reported around massive B-type forming stars, based on high-resolution observations of molecular and continuum tracers of disks \citep[e.g.,][]{Plambeck2016ApJ833219P,Ginsburg2018ApJ86119G,Beltran2016AARv246B,Cesaroni2014AA566A73C,chen2016ApJ823125C,Zhang2019ApJ87373Z,sanchez-monge2013AA552L10S,Beltran2014AA571A52B,Beltran2016AARv246B,Moscadelli2019AA622A206M}.
For O-type stars, this hypothesis can be directly confirmed by the detection of Keplerian disks around a forming O-type star. 
So far, there are seven detections of Keplerian disk candidates around O-type high-mass stars, as summarized by \citet{Goddi2020ApJ90525G}. Three out of the seven cases are located in the same sky region of this work and all of them are found to be associated with large-scale CO outflows. 
These are the O-type stars G17.64$+$0.16 \citep{Maud2018AA620A31M,Maud2019AA627L6M}, AFGL 4176 \citep{Johnston2015ApJ813L19J,Johnston2020AA634L11J}, and IRAS16547$-$4247 \citep{Zapata2019ApJ872176Z,Tanaka2020ApJ900L2T}, which are associated with ATLASGAL clumps AGAL017.637$+$00.154, AGAL308.917$+$00.122, and AGAL343.128$-$00.062, respectively. 
The disk-like structures in the three O-type protostars are found to be perpendicular to the CO outflows in G17.64$+$0.16  \citep{Maud2018AA620A31M} and AFGL 4176 \citep{Johnston2015ApJ813L19J}, and to the jet-axis orientation for IRAS16547$-$4247 \citep{Zapata2019ApJ872176Z}. 
The relative orientation of the disk-axis and outflow direction supports the interpretation these are driven by a central, massive protostar \citep{Kraus2010Natur466339K,Beltran2016AARv246B}.

There is mounting observational evidence showing that outflows and accretion disks are a common feature of high-mass star formation. 
The high detection rate of outflows reported here provides further indirect evidence of accretion disks in the formation of high-mass stars and indicate that they are not only common but are likely to be a ubiquitous feature of high-mass star formation, in other words, supporting the scenario that high-mass star formation is a scaled-up version of the mechanisms seen in low-mass star formation. However, this is almost certainly an oversimplification of what is undoubtedly a much more complicated process (i.e., \citealt{Goddi2020ApJ90525G}).

\section{Summary and conclusion}
\label{sect:summary_conclusion}

We conducted a CO outflow search covering an area of 68\,sq.\,degs and including 2052 ATLASGAL clumps using the simultaneously observed \coaa\ and \cobb\ lines from the SEDIGISM survey. 
1192 of these clumps show the signatures of high-velocity wings, indicative of outflows. Among the 1192 outflow candidates, 706 show bipolar wings and 486 show unipolar wings (256 blue wings and 230 red wings). This is the largest systematic sample of outflow candidates to date. We provide tables of the physical properties of the clumps associated with the outflows and the properties of the outflows themselves. We use these properties to  investigate the detection rates as a function of the evolutionary state of the clumps and physical properties, and the results are in agreement with Paper I. Our main findings are as follows:

\begin{enumerate}

\item 
The overall outflow detection rate is 58\%$\pm$1, and reduces to 34\%$\pm$1 for bipolar wings and 24\%$\pm$1 for unipolar wings, including 13\% and 11\% for unipolar blue and red wings, respectively. \\

\item
The outflow detection rate increases with increasing bolometric luminosities $\rm L_{bol}$, luminosity-to-mass ratios $\rm L_{bol}/M_{clump}$, and densities, in agreement with Paper I. 
\\
 
\item
Outflow detection rate increases as the clumps evolve, from the earliest quiescent (51\%), to protostellar (47\%), to YSO (57\%), and to MSF clumps (70\%), approaching $\sim$80\% for clumps associated with masers and \uchii\ regions. 
\\

\item 
Outflow detection rates are most closely related to the clump luminosity $\rm L_{bol}$. 
The increase in detection rate as a function of the evolutionary stages and the properties of clumps may be due to an increase in detectability rather than a lower fraction of outflow frequency. 
The low outflow detection rates for clumps in the earlier stages are possibly due to the  association with weak wings that are difficult to detect. 
The detection of outflows for the 70\,$\mu$m dark clumps suggests that the absence of 70\,$\mu$m emission is not a robust indicator of starless and/or pre-stellar cores.
 \\

\item The high detection rate from this large sample of clumps supports the scenario that high-mass star formation is a scaled-up version of low-mass star formation involving molecular outflows and accretion disks.\\

\end{enumerate}

In conclusion, outflows are common features in all evolutionary stages of massive star formation. 
Combining the large outflow sample in Paper\,I, we obtain the largest outflow sample so far, consisting of $\sim$1500 outflow-harboring clumps. 
This large and unbiased sample provides a statistically significant sample of interesting clumps in every evolutionary stage, especially for clumps (1) in the earliest evolutionary stages\,(i.e., 70$\mu m$ dark), and (2) with extremely high-velocity wings. 
Large-scale outflows can provide indirect evidence for the association of disk-like structures around massive protostars. 
High-resolution observations to map outflows on small scales for the outflow-harboring clumps are the next steps for investigation.   

\begin{acknowledgements}
We would like to thank the anonymous referee for the helpful comments and suggestions that have helped improve the clarity of this work.
H.B. acknowledges support from the European Research Council under the European Community's Horizon 2020 framework program (2014-2020) via the ERC Consolidator Grant "From Cloud to Star Formation (CSF)" (project number 648505). 
H.B. also acknowledges funding from the Deutsche Forschungsgemeinschaft (DFG) via the Collaborative Research Center (SFB 881) "The Milky Way System" (subproject B1). 
{This work was partially funded
by the Collaborative Research Council 956 "Conditions and impact
of star formation" (subproject A6), also funded by the DFG.}
{ADC acknowledges the support from the Royal Society University Research Fellowship (URF/R1/191609).}
L.B. acknowledges support from ANID BASAL project FB210003.
This publication is based on data acquired with the Atacama Pathfinder Experiment (APEX). 
APEX is a collaboration among the Max-Planck-Institut fur Radioastronomie, the European Southern Observatory, and the Onsala Space Observatory. 
The SEDIGISM survey includes projects 092.F-9315 and 193.C-0584, and the processed data products are available at https://sedigism.mpifr-bonn.mpg.de/index.html, which was constructed by James Urquhart and hosted by the Max Planck Institute for Radio Astronomy. 
The ATLASGAL project is a collaboration between the Max-Planck-Gesellschaft, the European Southern Observatory (ESO) and the Universidad de Chile, which includes projects 078.F-9040, 079.C-9501, 080.F-9701, 081.C-9501, 082.F-9701, 085.F-9505, 085.F-9526, 181.C-0885, with data available at https://www3.mpifr-bonn.mpg.de/div/atlasgal/.
This research made use of Astropy\footnote{http://www.astropy.org}, a community-developed core Python package for Astronomy \citep{Astropy2013AA33A,Astropy2018AJ}. 
This research has made use of the SIMBAD database and the VizieR catalogue, operated at CDS, Strasbourg, France. 
This document was prepared using the Overleaf web application available at www.overleaf.com.

\end{acknowledgements}

\bibliographystyle{aa}
\bibliography{ref}


\vspace{0.5cm}
{\it \small
\noindent $^{1}$ Max Planck Institute for Radio Astronomy, Auf dem H\"ugel 69, 53121, Bonn, Germany \\ 
$^{2}$ Centre for Astrophysics and Planetary Science, University of Kent, Canterbury, CT2 7NH, UK \\ 
$^{3}$ School of Physics and Astronomy, University of Leeds, Leeds, LS2 9JT, UK \\   
$^{4}$ School of Physics and Astronomy, Cardiff University, Cardiff CF24 3AA, UK \\  
$^{5}$ Leibniz-Institut f\"ur Astrophysik Potsdam (AIP), An der Sternwarte 16, 14482 Potsdam, Germany \\  
$^{6}$ Laboratoire d'astrophysique de Bordeaux, Univ. Bordeaux, CNRS, B18N, all\'ee Geoffroy Saint-Hilaire, 33615 Pessac, France \\  
$^{7}$ Astrophysics Research Institute, Liverpool John Moores University, 146 Brownlow Hill, Liverpool, L3 5RF, UK \\ 
$^{8}$ Space Science Institute, 4765 Walnut Street Suite B, Boulder, CO 80301, USA \\ 
$^{9}$ Istituto di Astrofisica e Planetologia Spaziali, INAF, via Fosso del Cavaliere 100, I-00133 Roma, Italy \\ 
$^{10}$ Departamento de Astronom\'ia, Universidad de Chile, Casilla 36-D, Santiago, Chile \\ 
$^{11}$ I. Physikalisches Institut, Universit\"at zu K\"oln, Z\"ulpicher Str. 77, D-50937 K\"oln, Germany \\ 
$^{12}$Department of Astronomy, University of Florida, PO Box 112055, USA \\ 
$^{13}$Korea Astronomy and Space Science Institute, 776 Daedeok-daero, 34055 Daejeon, Republic of Korea \\ 
$^{14}$Observatorio Astrofisico di Arcetri, Largo Enrico Fermi 5, I-50125 Firenze, Italy	\\  
$^{15}$Max-Planck-Institut f\"ur Astronomie, K\"onigstuhl 17, D-69117 Heidelberg, Germany \\
$^{16}$ School of Science and Technology, University of New England, Armidale NSW 2351, Australia \\
} 


\end{document}